  \documentclass{aa}
\usepackage{graphicx}
\usepackage{txfonts}
\usepackage{natbib}
\usepackage{longtable}
\bibpunct{(}{)}{;}{a}{}{,}   

\begin{document}
   \title{Volume-limited radio survey of ultracool dwarfs}

\author{A.~Antonova\inst{1}
\and G.~Hallinan\inst{2,3}
\and J.~G.~Doyle\inst{4}
\and S. Yu\inst{4}
\and A. Kuznetsov\inst{4,5}
\and Y. Metodieva\inst{1}
\and A.~Golden\inst{6,7}
\and K.~L.~Cruz\inst{8,9}
}

   \offprints{A. Antonova, tony@phys.uni-sofia.bg}

   \institute{Department of Astronomy, St. Kliment Ohridski University of Sofia, 5 James
   Bourchier Blvd., 1164 Sofia, Bulgaria
  \and National Radio Astronomy Observatory, 520 Edgemont Road, Charlottesville, VA 22903, USA
  \and Department of Astronomy, University of California, Berkeley, CA 94720, USA
  \and Armagh Observatory, College Hill, Armagh BT61 9DG, N. Ireland
  \and Institute of Solar-Terrestrial Physics, Irkutsk 664033, Russia
  \and Centre for Astronomy, National University of Ireland, Galway, Ireland
  \and Price Center, Albert Einstein College of Medicine, Yeshiva University, Bronx, NY 10461, USA
  \and Department of Physics and Astronomy, Hunter College, City University of New York, 10065, New York, NY, USA
  \and Department of Astrophysics, American Museum of Natural History, 10024, New York, NY, USA
}

   \date{Received --||-- / Accepted by A\&A, 19 Nov 2012}
\titlerunning{A Volume Limited Radio Survey of Ultracool Dwarfs}


  \abstract
  {}
{We aim to increase the sample of ultracool dwarfs studied in the radio
domain to allow a more statistically significant understanding of the
physical conditions associated with these magnetically active objects. }
{We conducted a volume-limited survey at 4.9 GHz of 32 nearby ultracool dwarfs
with spectral types covering the range M7 -- T8. 
A statistical analysis was performed on the combined data from the
present survey and previous radio observations of ultracool dwarfs. }
{Whilst no radio emission was detected from any of the targets, significant upper limits were placed on
the radio luminosities that are below the luminosities of previously detected
ultracool dwarfs. Combining our results with those from the literature
gives a detection rate for dwarfs in the spectral range M7 -- L3.5
of $\sim 9\%$. In comparison, only one dwarf later than L3.5 is detected
in 53 observations. We report the observed detection rate as a function of
spectral type, and the number distribution of the dwarfs as a
function of spectral type and rotation velocity. } 
{The radio observations to date point to a drop in the detection rate
toward the ultracool dwarfs. However, the emission levels of detected ultracool dwarfs
are comparable to those of earlier type active M dwarfs, which may imply that a
mildly relativistic electron beam or a strong magnetic field can
exist in ultracool dwarfs. Fast rotation may be a sufficient condition to produce magnetic fields  
strengths of several hundreds Gauss to several kilo Gauss, as suggested by the data for the active ultracool dwarfs with known
rotation rates. A possible reason for the non-detection of radio emission from some dwarfs is that maybe the
centrifugal acceleration mechanism in these dwarfs is weak (due to a
low rotation rate) and thus cannot provide the necessary density
and/or energy of accelerated electrons. An alternative explanation
could be long-term variability, as is the case for several ultracool
dwarfs whose radio emission varies considerably over long periods
with emission levels dropping below the detection limit in some
instances.}

   \maketitle

\section{Introduction} \label{sec:intro}
In the past few years, a number of very low mass stars and brown
dwarfs (collectively termed ultracool dwarfs) have been confirmed
as a new class of radio active objects. A surprising feature of
these observations is the detection of periodic pulses of 100$\%$
circularly polarized emission \citep{hallinan06, hallinan07,
hallinan08, berger09, Doyle2010, mclean+etal2011}. These periodic
pulses have been confirmed to be produced by the electron cyclotron
maser (ECM) instability, the same mechanism that is known to produce the
planetary radio emission at kHz and MHz frequencies
\citep{treumann06}, but it requires much more powerful kilogauss
magnetic fields. Unpolarized and seemingly quiescent radio emission
was present in the observations of all detected dwarfs and has
alternatively been attributed to gyrosynchrotron emission and the
ECM emission (see the above references).

Although much progress has been made in understanding the nature of the pulsed radio emission from
these dwarfs by using its diagnostic potential, it remains unclear which characteristics
distinguish radio `active' from radio inactive dwarfs. Possible physical characteristics include mass,
temperature, activity, and rotation rate. Intriguingly, all active dwarfs have been found
to have high $v \sin i$ values. This suggests two possibilities, a dependence on rotation velocity,
or a dependence on inclination angle. The former suggests that slower rotators have weaker dynamo
action and hence weaker magnetic fields of insufficient strength to produce detectable radio emission
at the required frequency. The second scenario implies a dependence on inclination angle, i.e.,
a geometrical selection effect is associated with the highly beamed radio emission. A recent
case study of three pulsing ultracool dwarfs does indeed confirm that all three have very
high values of inclination (greater than 65 degrees). However, all three dwarfs are also confirmed
to be very rapid rotators with periods of rotation $\leq$ 3 hours \citep{hallinan08}.

Clarifying the relationship between $v \sin i$ and radio activity is imperative. If a dependence
on rapid rotation underlies the observed correlation between radio luminosity and $v \sin i$, this implies
that the rotation-activity relationship, which is well-established for main-sequence stars, extends into the
substellar regime. If, on the other hand, the correlation between radio luminosity and $v \sin i$
indicates a geometrical selection effect, this implies that very strong magnetic fields (kG) are
ubiquitous in the substellar regime, independent of rotation rate. Thus, to clarify the
relationship between $v \sin i$ and radio activity, a larger more statistically significant sample of
pulsing dwarfs must be established.

The ECM emission has proved a vital diagnostic tool for remote sensing of the magnetic
field strength and topologies of ultracool dwarfs. The ECM emission is generated at the electron
cyclotron frequency denoted by $\nu_{\rm c} \approx 2.8 \times 10^{6}~B$ Hz (if the electrons are not
relativistic), enabling measurement of the magnetic field strength in the source region of the pulsed emission.
This led to the realization of kG magnetic fields in late-M dwarfs \citep{hallinan07}
and subsequently the first confirmation of kG magnetic fields for an L dwarf, the latter establishing
strong magnetic dynamo action out to spectral type L3.5 \citep{hallinan08}.

Very recently, \citet{Route12} reported the first detection of radio emission from a dwarf of spectral type later than L3.5 - the T6.5 brown dwarf
2MASS J1047539+212423. They have detected circularly polarized bursts at 4.75 GHz with the Arecibo telescope and 
invoked the ECM mechanism as the most likely source of the emission. This latest observation confirms that detecting ECM 
emission remains the most promising method to measure magnetic field strengths in cooler late-type L and T
dwarfs. \citet{reiners+basri07} measured the magnetic field strengths of a number of late-M
dwarfs through the measurement of the Zeeman broadening of the magnetically sensitive Wing-Ford FeH band.
However, this technique encounters difficulties when applied to L and T dwarfs due to the heavy saturation
of the FeH lines. Therefore, the continued search for radio emission from late-L and T dwarfs is essential
to the efforts to diagnose the strength of magnetic fields in these objects.

To address the above questions we conducted a volume-limited survey of 32 ultracool dwarfs of spectral types M7 to T8.

\begin{table*}
\begin{center}
\caption{Sample of 32 dwarfs, their properties, and results from the present survey.} \label{tab:table1}
{\scriptsize
 \begin{tabular}{lllllllcc}
\hline
Name                       & Other name             & Sp. T.  & d        &  $v \sin i$  & L$_{\mathrm{bol}}$ &(L$_{\mathrm{H\alpha}}$ / L$_{\mathrm{bol}}$) & $F_{\mathrm{(4.9~GHz)}}$ & L$_{\nu,4.9}$ \\
                          &                         &         &(pc)      & (km s$^{-1}$)&  (L$_{\odot}$) &  & ($10^{-5}$Jy)   &  (erg~s$^{-1}$~Hz$^{-1}$) \\
\hline
2MASS J10481258-1120082  &  GJ 3622         &  M7      &   4.5   &  3.0& - 3.16 & - 4.63     & $<$ 6.3  & $<$ 1.53 $\times 10^{12}$   \\
2MASS J17571539+7042011  &  LP 44-162           &  M7.5    &   12.5  &  33 & - 3.48 & - 5.01    & $<$ 8.1  & $<$ 1.51 $\times 10^{13}$   \\
2MASS J11554286-2224586  &  LP 851-346          &  M7.5    &   9.7   &  33 &  --    & - 4.58    & $<$ 6.6  & $<$ 7.43 $\times 10^{12}$   \\
2MASS J05395200-0059019  &  SDSS J053951.99-005902.0  &  M7.5    &   3.84  &  -- &  --    &  --    & $<$ 6.0  & $<$ 1.06 $\times 10^{12}$   \\
2MASS J12505265-2121136  &  DENIS-P J125052.6-212113  &  M7.5    &   11.1  &  -- &  --    &  --    & $<$ 6.6  & $<$ 9.73 $\times 10^{12}$   \\
2MASS J04351455-1414468  &  --                      &  M8      &   14    &  -- &  --    &  --      & $<$ 6.0  & $<$ 1.41 $\times 10^{13}$   \\
2MASS J02150802-3040011  &  LHS 1367, LP885-35      &  M8      &   12.37 &  -- & - 3.55 &  --      & $<$ 7.5  & $<$ 1.37 $\times 10^{13}$   \\
2MASS J05392474+4038437  &  LSR J0539+4038          &  M8      &   10    &  -- &  --    &  --      & $<$ 6.3  & $<$ 7.54 $\times 10^{12}$   \\
2MASS J18261131+3014201  &  LSR J1826+3014          &  M8.5    &   13.9  &  -- &  --    &  --      & $<$ 8.7  & $<$ 2.01 $\times 10^{13}$   \\
2MASS J14284323+3310391  &  GJ 3849, LHS 2924       &  M9      &   11.8  &  10 & - 3.62 &  - 4.7     & $<$ 6.3  & $<$ 1.04 $\times 10^{13}$   \\
2MASS J17312974+2721233  &  LSPM J1731+2721          &  L0     &   11.8  &  15 & - 3.74 &  - 4.6     & $<$ 5.7  & $<$ 9.50 $\times 10^{12}$   \\
2MASS J09211410-2104446  &  DENIS-P J092114.1-210445 &  L2     &   12    &  15 & - 4.01 & $<$ - 6.42& $<$ 7.2  & $<$ 1.14 $\times 10^{13}$   \\
2MASS J08283419-1309198  &  DENIS-P J082834.3-130919  &  L2      &   11.6  &  33  &   --   & - 5.68   & $<$ 6.3  & $<$ 1.01 $\times 10^{13}$   \\
2MASSI J0700366+315726   &   --                     &  L3.5+L6 &   12.2  &  41 &- 3.96  &  --      & $<$ 4.2  & $<$ 8.01 $\times 10^{12}$   \\
2MASS J05002100+0330501  &   --                     &  L4      &   13.03 &  -- &- 4.26  &  --       & $<$ 5.1  & $<$ 1.04 $\times 10^{13}$   \\
2MASS J04351455-1414468  &   --                     &  L4.5    &   9.8   &  -- &- 4.12  &  --       & $<$ 4.2  & $<$ 4.83 $\times 10^{12}$   \\
2MASS J03552337+1133437  &   --                     &  L5      &   12.6  &  10&- 4.03  &  --       & $<$ 4.5  & $<$ 8.55 $\times 10^{12}$   \\
2MASS J05395200-0059019  &SDSS J053951.99-005902.0  &  L5      &   13.1  &  34 &- 4.2   &  --       & $<$ 4.8  & $<$ 9.86 $\times 10^{12}$   \\
2MASS J17502484-0016151 &    --                     &  L5.5    &   8     &  -- & --     &  --       & $<$ 4.3  & $<$ 4.82 $\times 10^{12}$   \\
2MASS J21481633+4003594 &    --                     &  L6.5    &   7.9   &  -- &- 4.07  &  --       & $<$ 6.3  & $<$ 4.70 $\times 10^{12}$   \\
2MASS J09083803+5032088 &    --                     &  L7      &   15.9  &  31 & --     &  --       & $<$ 11.1 & $<$ 3.36 $\times 10^{13}$   \\
2MASS J08575849+5708514 & SDSS J085758.45+570851.4  &  L8      &   11    &  -- & --     &  --       & $<$ 5.1  & $<$ 7.38 $\times 10^{12}$   \\
2MASS J02572581-3105523 & DENIS-P J025725.7-310552  &  L8      &   9.6   &  -- &- 4.82  &  --       & $<$ 9.3  & $<$ 1.03 $\times 10^{13}$   \\
2MASS J08300825+4828482 & SDSS J083008.12+482847.4 &  L9      &   13.0  &  -- &- 4.58  &  --       & $<$ 8.7  & $<$ 1.78 $\times 10^{13}$   \\
2MASS J01365662+0933473  &  SIMP J013656.5+093347.3 &  T2.5    &   6.4   &  -- & --     &  --       & $<$ 7.5  & $<$ 3.68 $\times 10^{12}$   \\
             &              &          &   8.27  &     &        &       &          & $<$ 6.14 $\times 10^{13}$   \\
2MASS J15031961+2525196  &      --          &  T5.5    &   7     &  36 & --     &  --       & $<$ 9.0    & $<$ 5.28 $\times 10^{12}$   \\
2MASS J09373487+2931409  &      --          &  T6      &   6.14  &  -- &- 5.28  &  --       & $<$ 6.6  & $<$ 2.98 $\times 10^{12}$   \\
2MASS J21543318+5942187  &      --          &  T6      &   18.8  &  -- & --     &  --       & $<$ 6.0  & $<$ 2.54 $\times 10^{13}$   \\
2MASS J07271824+1710012  &      --          &  T7      &   9.09  &  -- &- 5.26  & $<$ - 5.4  & $<$ 5.4  & $<$ 5.34 $\times 10^{12}$   \\
2MASS J11145133-2618235  &      --          &  T7.5    &   7     &  -- & --     &  --       & $<$ 6    & $<$ 3.52 $\times 10^{12}$   \\
             &              &          &   10    &     &        &       &          & $<$ 7.18 $\times 10^{12}$   \\
2MASS J09393548-2448279  &      --          &  T8      &   8.7   &  -- &- 5.69  &   --      & $<$ 5.4  & $<$ 4.89 $\times 10^{12}$   \\
             &              &          &   5.34  &     &        &      &          & $<$ 1.84 $\times 10^{12}$   \\
2MASS J14571496-2121477  &  GL570D  &  T8      &   5.88  &  30 & - 5.52 &  --      & $<$ 5.4  & $<$ 2.23 $\times 10^{12}$ \\
\hline
\end{tabular}
}
\parbox{160mm}{
The columns are (left to right): name of the object; other name; spectral type; distance; projected rotational velocity; bolometric luminosity; H$\alpha$ luminosity; radio flux at 4.9 GHz; radio luminosity at 4.9 GHz. Spectral types, distances and $v~\sin~i$ measurements taken from: SIMBAD, \citet{bernat+etal2010,goldman+etal2010,lee+etal2010,reiners+basri2010a,Seifahrt2010,smart+etal2010,delBurgo+etal09,faherty+etal09,leggett+etal09,stephens+etal09,looper+etal08,reid+etal08,reiners+basri08,blake+etal07,kendall+etal07,looper+etal07,reiners+basri07,schmidt+etal07,artigau+etal06,henry+etal06,morales+etal06,reid+etal06,zapos+etal06,crifo+etal05,cushing+etal05,lodieu05,tinney+etal05,golim+etal04,vrba+etal04,burgasser+etal03,cruz+etal03,lepine+etal03,lepine+etal02}
}
\end{center}
\end{table*}
\section{Observations and data reduction}
\label{sec:observ}
The target list was compiled using the 2MU2 20 pc volume limited catalog of \citet{cruz_etal2007}.
We selected field dwarfs that had no previous observations at radio frequencies and whose declination
is $>$~-30$^{\circ}$ (the latter ensuring that the observing time is sufficiently long for achieving the
goals described below). This yielded 14 dwarfs in the spectral range M7 - L3.5, with a
view to expanding the current sample of detected pulsating ultracool dwarfs. We also selected 18 dwarfs of types L4 to T8 with the aim to
detect radio emission from a source later than L4. Their properties are listed in Table \ref{tab:table1}.

The observations were conducted with the NRAO Very Large Array\footnote{The National Radio Astronomy
Observatory is a facility of the National Science Foundation operated under cooperative agreement by
Associated Universities, Inc.} in the period 09 August - 09 September 2009 using the standard
continuum mode with 2 $\times$ 50 MHz contiguous bands. The observing frequency was chosen to be 4.9
GHz, because any ultracool dwarf observed at both 4.9 GHz and 8.5 GHz has been found to produce quiescent
emission at 4.9 GHz of the same luminosity or higher than that detected at 8.5 GHz. Furthermore, in
the case of the ECM emission, an upper cut-off frequency is expected for the radio emission that is
dependent on the maximum magnetic field of the dwarf. A survey at 4.9 GHz thus allows for
a lower magnetic field strength, which additionally increases the chance of
detection. For each source, a suitable phase calibrator was chosen from the VLA calibrator
manual while the flux density scale was determined using the calibrators 3C48, 3C138, and 3C286.

We observed each target in the survey for a duration determined by its distance, thereby ensuring a
common upper limit on the luminosity of the radio emission from each dwarf. For the dwarfs spanning
the spectral range M7 - L3.5, the duration was chosen to match the one required to place an upper limit
on the radio luminosity of $\sim 10^{13}~$ergs s$^{-1}$~Hz$^{-1}$, i.e., lower than each of the previously
detected pulsing ultracool dwarfs. Considering that no detection has previously been reported for
late L and T dwarfs, we chose to observe each of the later type ($\geq$ L4) dwarfs for the duration required to
place an upper limit on radio luminosity of $\sim 5 \times 10^{12}$~ergs s$^{-1}$~Hz$^{-1}$, i.e.,
to partially allow for any drop off in emission that might occur with
later spectral type. All targets were observed in a dynamic scheduling mode with a time resolution of
3.3 sec. This mode was therefore sufficient to detect short radio bursts, in addition to any quiescent
emission.

Data reduction was carried out with the Astronomical Image Processing System (AIPS) software package
using standard routines. The visibility data were inspected for quality both before and after the
calibration procedures, and noisy points were removed. To make the maps we used the task IMAGR and
then CLEANed the region around each source in the individual fields.

\section{Results}
\label{sec:results}

We found no sources at (or near) the positions of the studied dwarfs in the respective maps, i.e., we
did not detect radio emission from any of the observed dwarfs. As discussed in Section \ref{sec:observ},
in the case of non-detections we aimed to obtain data with a sensitivity that would allow us to place
upper limits of L$_{\nu} \leq 10^{13}~$ergs s$^{-1}$~Hz$^{-1}$ for the dwarfs in the range M7 - L3.5 and L$_{\nu} \leq 5 \times 10^{12}$~ergs s$^{-1}$~Hz$^{-1}$
for the later type dwarfs (L4 - T8).  We find that for all dwarfs in the sample whose distances are
within $\sim$ 12 pc, the detection limits are at or below these values. For the most distant dwarfs, however,
the limits were not reached due to insufficient sensitivity. The $3\sigma$ upper limits on the flux and the respective radio luminosity limits are
shown in Table \ref{tab:table1}.

 \begin{figure*}[htp!]
 \centering
 \resizebox{\hsize}{!}{
 \includegraphics[angle=0]{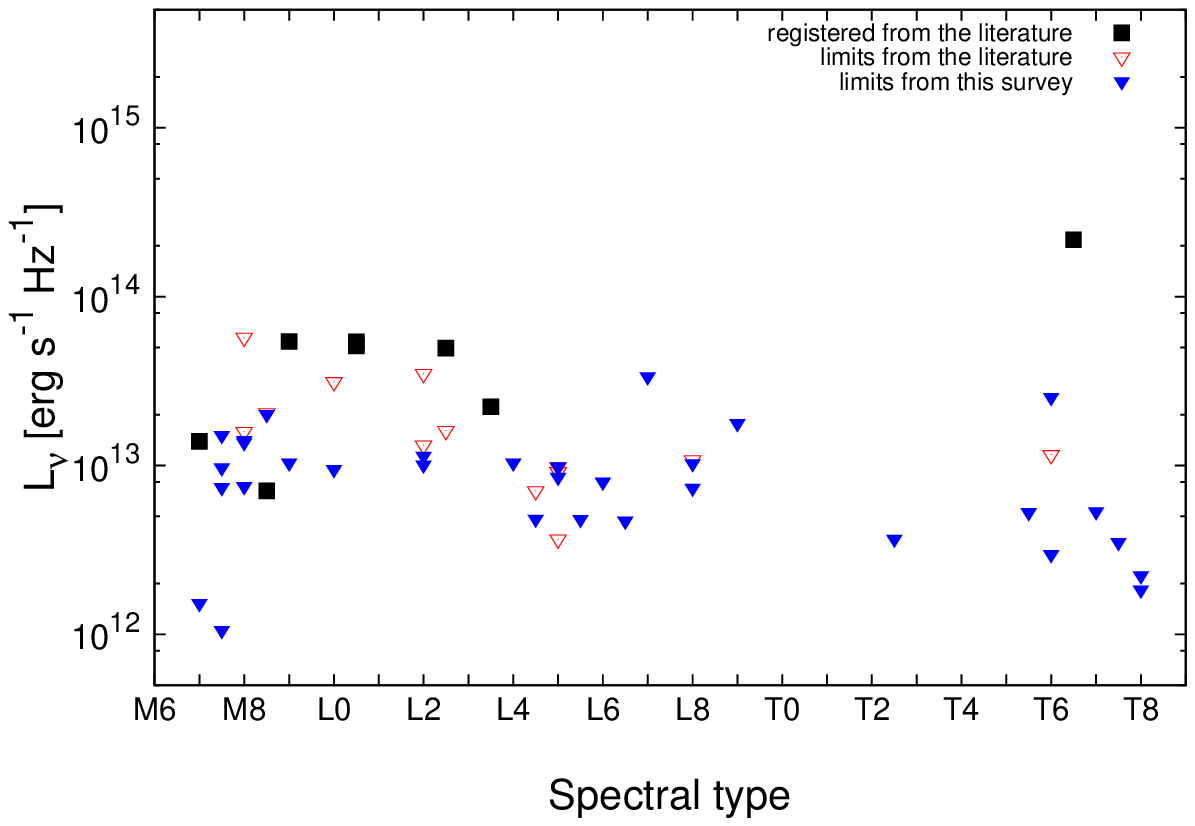}
 }
 \caption[]{Radio luminosity - spectral type plot for all ultracool dwarfs observed at 4.9 GHz. Filled 
 triangles show the upper limits on radio luminosities of the dwarfs from the present survey, open triangles 
 mark upper limits from the literature, and filled squares represent radio luminosities of detected dwarfs 
 from the literature \citep[][and references therein]{berger06, burgasser05, antonova07, antonova08, Route12}.}
 \label{fig:figure1}
 \end{figure*}

We plot our results together with those from previous studies of UCDs at 4.9 GHz in Figure \ref{fig:figure1}. 
This work increases the number of studied objects by more than 150$\%$ and lowers the 
upper limits on the radio luminosity. There is only one upper limit whose value is above the value of detected 
emission for a given spectral type - that for the M8.5 dwarf LSR J1826+3014, which is one of the most distant 
dwarfs in the survey (d = 13.9 pc). 

\citet{mclean+etal2012} recently reported on a 8.5 GHz radio survey of 100 M and L dwarfs spanning the
range M4 to L4 (76 of which are ultracool dwarfs) and the detection of three new radio-emitting late-type
dwarfs. Comparing the target list of their observation to ours, we found seven dwarfs in common, none of
which was detected at the higher frequency either. We list these dwarfs and their flux and luminosity
upper limits in the respective frequencies in Table \ref{tab:table2}. We have calculated the upper
limit on the 8.5 GHz radio luminosity using the distances listed in Table \ref{tab:table2} and the
fluxes reported by \citet{mclean+etal2012}. Despite the small differences in the 3$\sigma$ limits on
the radio flux  between the two surveys, we find that the limits in radio luminosities agree well.

In addition, \citet{mclean+etal2012} compiled a list of all previous radio observations of
late-type dwarfs. Combining their list with the present survey raises the total number of studied
ultracool dwarfs to 193, with 13 detected to posses radio emission (see Figure \ref{fig:figure2}).
Although this might suggest a detection rate of $\sim 6\%$ in the ultracool dwarf regime, a more detailed 
analysis indicates otherwise. All but one of the detected ultracool dwarfs span the spectral type range M7 - L3.5.
Therefore, the detection rate over this spectral range is 12 confirmed sources from 134
observations, i.e., $\sim 9\%$. In comparison, only one dwarf later than L3.5 was detected in 53 observations.

\begin{table*}
\begin{center}
\caption{Upper limits on the radio flux and radio luminosity at 4.9 GHz (present survey) and 8.5 
GHz \citep{mclean+etal2012} of the seven dwarfs that are present in both surveys. We used the distances listed in Table\ref{tab:table1} to calculate L$_{\nu, 8.5}$.} \label{tab:table2}
 \begin{tabular}{lllllllcc}
\hline
Name                     & Other name             & $F_{\mathrm{(4.9~GHz)}}$ & L$_{\nu, 4.9}$              &$F_{\mathrm{(8.5~GHz)}}$ & L$_{\nu, 8.5}$\\
                         &                        & ($10^{-5}$Jy)        &  (erg~s$^{-1}$~Hz$^{-1}$)  & ($10^{-5}$Jy)       &  (erg~s$^{-1}$~Hz$^{-1}$) \\
\hline
2MASS J10481258-1120082  &  GJ 3622         &$<$ 6.3 & $<$ 1.53 $\times 10^{12}$   & $<$ 9.6  &  $<$ 2.33 $\times 10^{12}$\\
2MASS J17571539+7042011 &  LP 44-162            &$<$ 8.1 & $<$ 1.51 $\times 10^{13}$   & $<$ 11.7 &  $<$ 2.19 $\times 10^{13}$ \\
2MASS J11554286-2224586  &  LP 851-346          &$<$ 6.6 & $<$ 7.43 $\times 10^{12}$   & $<$ 9.0  &  $<$ 1.01 $\times 10^{13}$\\
2MASS J12505265-2121136          &   &$<$ 6.6 & $<$ 9.73 $\times 10^{12}$   & $<$ 7.2  &  $<$ 1.06 $\times 10^{13}$\\
2MASS J09211410-2104446 &   &$<$ 7.2 & $<$ 1.14 $\times 10^{13}$   & $<$ 7.5  &  $<$ 1.29 $\times 10^{13}$\\
2MASS J08283419-1309198 &   &$<$ 6.3 & $<$ 1.01 $\times 10^{13}$   & $<$ 6.6  &  $<$ 1.06 $\times 10^{13}$\\
2MASSI J0700366+315726   &   &$<$ 4.2 & $<$ 8.01 $\times 10^{12}$   & $<$ 7.8  &  $<$ 1.39 $\times 10^{13}$\\

\hline
\end{tabular}
\end{center}
\end{table*}

\begin{figure*}[htp!]
\hspace*{-0.8cm}
\centering
\includegraphics[width=18cm,clip,angle=00]{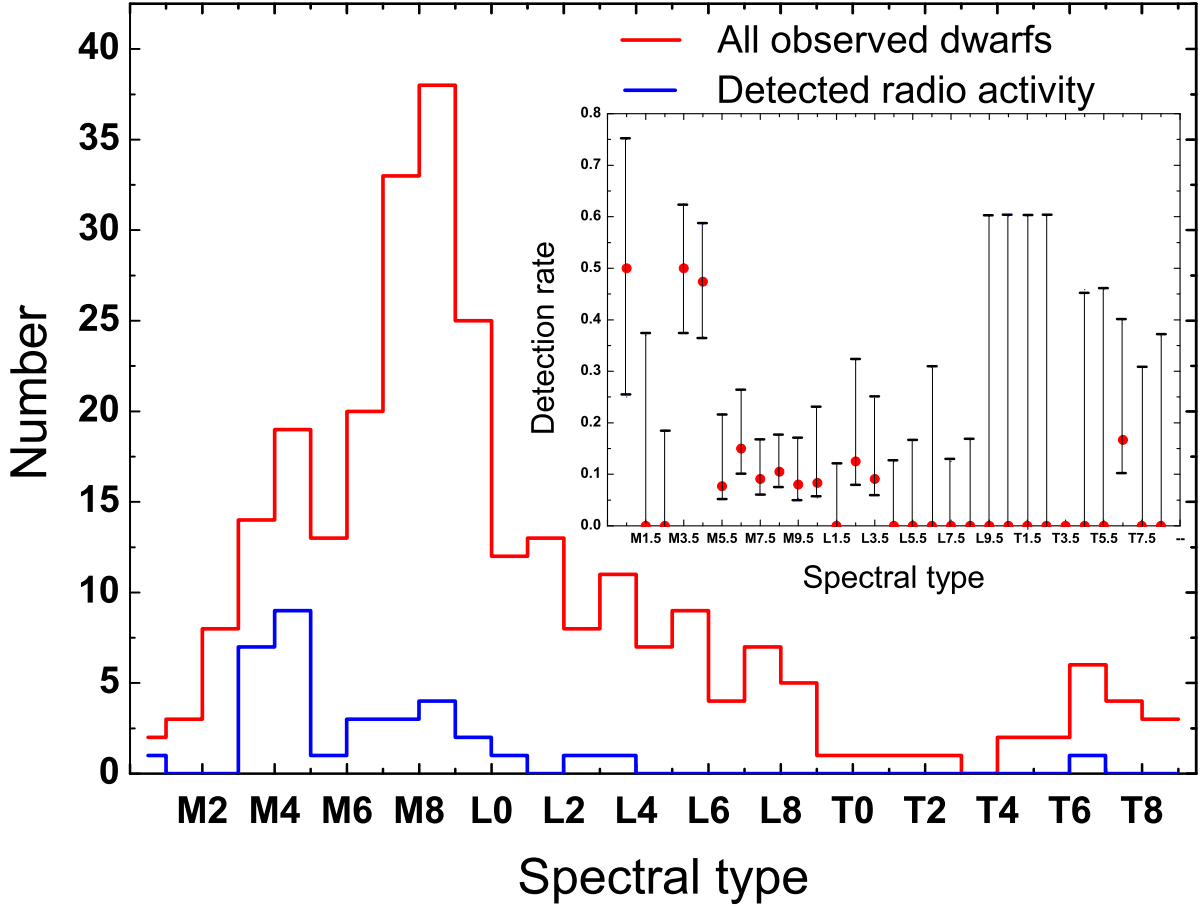}
\caption{Number distribution of observed dwarfs (red line) and dwarfs with detected radio signal (blue line) with respect to
their spectral type. The insert panel shows the detection rate as a function of spectral type, defined by $N_{\rm r}/N$, where
$N_{\rm r}$ is the number of dwarfs with detected radio signal in a certain spectral type range; $N$ is the number of observed dwarfs
in the same spectral type range. In the spectral type range of M0-T8, the total number of observed dwarfs is 273, while the number
of dwarfs with detected radio signal is 34. The zero detection rate around spectral type M2 is most likely 
due to the limited number of observations. 
Also shown are the uncertainties of the detection rates, estimated by adopting the binomial distribution at 1$\sigma$ 
level for the Gaussian distribution (see \S\,\ref{sec:results}). }
\label{fig_sptdis}
\end{figure*}

\begin{figure*}[htp!]
\hspace*{-0.8cm}
\centering
\includegraphics[width=18cm,clip,angle=00]{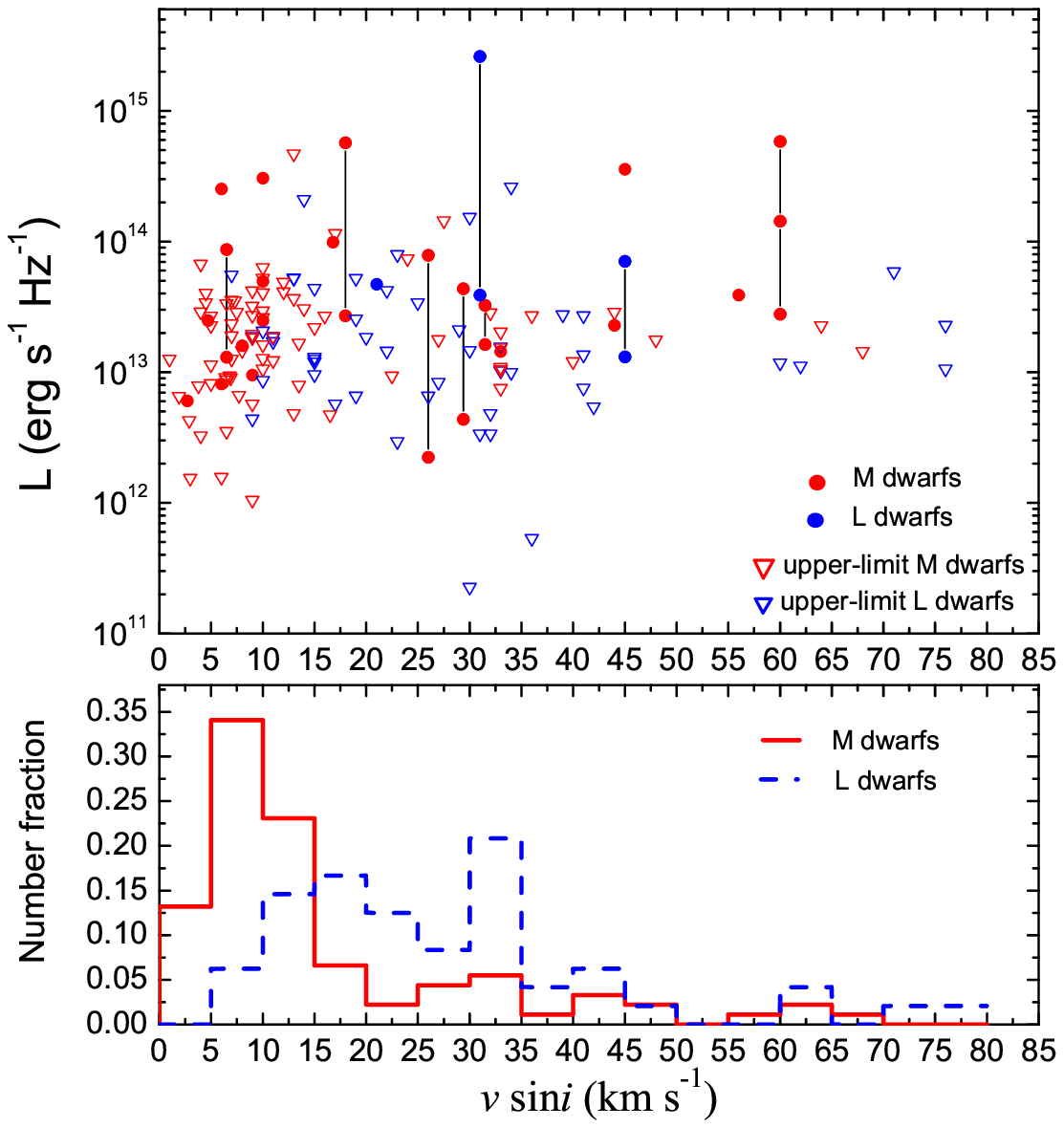}
\caption{Top panel: Radio luminosity as a function of rotation velocity for the UCDs observed in the radio domain
(mainly at 4.9 and 8.5 GHz).
Red solid circles are for M dwarfs, the blue solid circles represent L dwarfs. The solid circles connected by black lines
denote different measurements for the same target. Red open triangles denote the upper-limit of the luminosity of M dwarfs;
blue open triangles stand for the upper-limit of the luminosity of L dwarfs. The data are mainly taken from Table\,\ref{tab:table1}, \citet{mclean+etal2012}, and references therein. Bottom panel: the number
distribution of the observed dwarfs as a function of rotation velocity. The red solid line is for M dwarfs (normalized by the total number 91),
the blue dashed line is for L dwarfs (normalized by the total number 48). }
\label{fig_vsiniv}
\end{figure*}

In Fig.\,\ref{fig_sptdis}, we plot the number distribution of cool dwarfs later than M0 (red line) observed in the radio domain compared 
to those with a detected radio signal (blue line)
with respect to their spectral type. We include earlier type M dwarfs to show the overall trend in detection rate. The insert panel in the 
figure shows the detection rate as a function of spectral type, defined by $N_{\rm r}/N$, where $N_{\rm r}$ is the number of dwarfs with 
detected radio signal in a certain spectral type range; $N$ is the number
of observed dwarfs in the same spectral type range. To estimate the uncertainty of the detection rate as a function of spectral
type, we adopted the binomial distribution with the limits equivalent to
$1\sigma$ limits for a Gaussian distribution (see, e.g., the article of
\citet{burgasser+etal2003a}, where this model was applied to the statistics of
T-dwarf binaries). There is a drop in detection rate from $\sim 50\%$ at spectral type M4 to less than $15\%$ for 
later spectral types. The drop may be due to the small sample of observed UCDs, especially those later than L0. 
Alternatively, the drop may have physical implications. We discuss these possible implications in \S\,\ref{sec_mechanisms}. 

The top panel in Fig.\,\ref{fig_vsiniv} shows the observed radio luminosity as a function of rotation velocity for the M and L
dwarfs (most of them are observed by the VLA, mainly at 4.9 and 8.5 GHz). Solid circles are for the M and L dwarfs
with detected strong radio signal, whilst open triangles represent the upper-limit of the luminosity of some M and
L dwarfs. The relation between the luminosity and $v\sin i$ in this figure is not very clear. It seems that the number
of M dwarfs possessing strong radio emission is more than the number of L dwarfs. Even for one individual dwarf, the
detected radio luminosity can vary over a wide range (about 1 - 2 orders of magnitude) in a short observation period,
which may suggest that the radio emission is strongly associated with its generation mechanism and may not be directly correlated with rotation rate.

The bottom panel in Fig.\,\ref{fig_vsiniv} illustrates the number distribution of all observed M (red line) and L (blue line)
dwarfs as a function of rotation velocity ($v\sin i$). The number distribution of M dwarfs decreases with respect to
$v\sin i$, perhaps caused by a stellar-wind-assisted magnetic braking. More observations are needed to exclude
this selection effect. Also, since the majority of the data represent upper limits, no statistically significant conclusions 
can be inferred at present.

\begin{figure*}
 \centering
 \resizebox{\hsize}{!}{
 \includegraphics[angle=0]{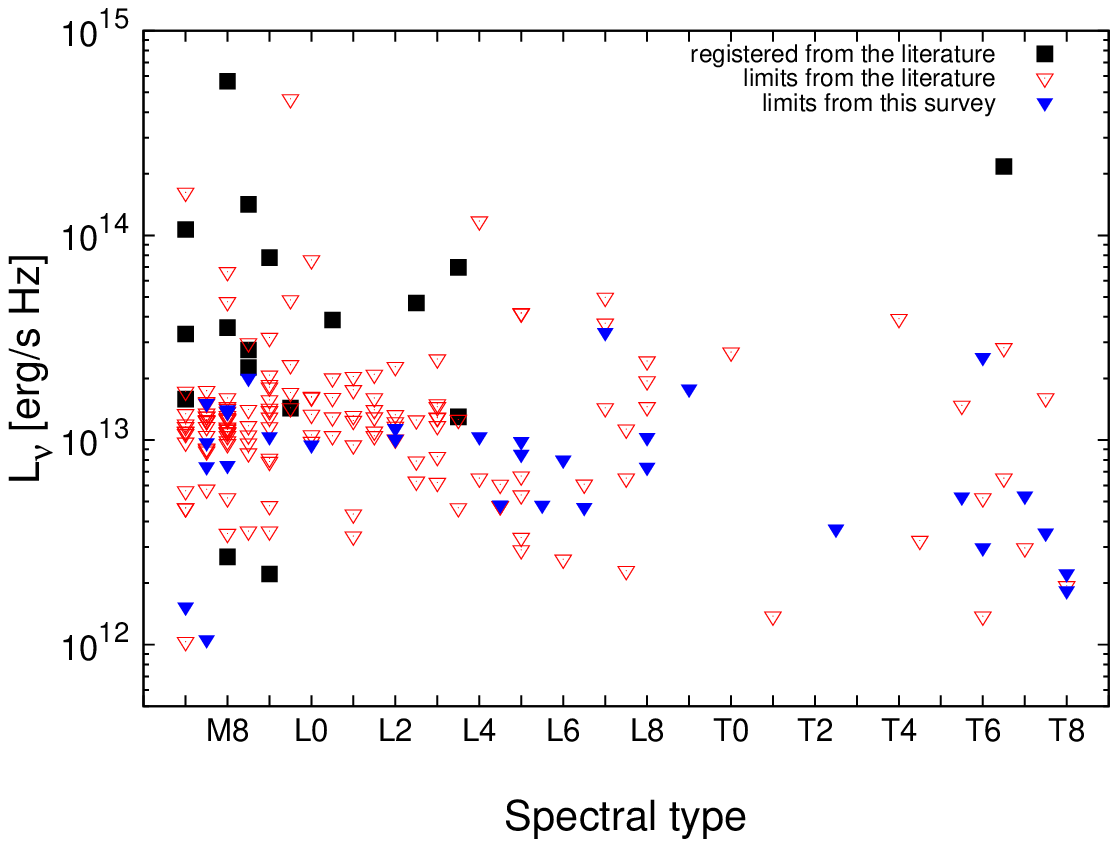}
 }
 \caption[]{Radio luminosity - spectral type plot for all ultracool dwarfs  observed at radio frequencies. 
 Filled triangles show the upper limits on radio luminosities of the dwarfs from the present survey. Open triangles 
 mark upper limits from the literature and filled squares represent radio luminosities of detected dwarfs from the 
 literature \citep[][and references therein]{mclean+etal2012}. }
 \label{fig:figure2}
 \end{figure*}

\section{Discussion}
\label{sec:discussion}

In this section, we discuss the physical implication of our survey, including the influence of the
radiation mechanism,  the geometric effect, and the rotation of the dwarfs. A brief discussion on two 
other activity characteristics - X-ray and H$\alpha$ emission from the radio observed sample - is also included in Appendix \ref{appA}.

\subsection{Radiation mechanisms and geometric effect}
\label{sec_mechanisms}

For 12 of the 13 radio active dwarfs a quiescent component of the radio emission was also detected. The 
T6.5 dwarf 2MASS J1047539+212423 was observed with the Arecibo single-dish, which has a low sensitivity to 
unpolarized quiescent emission, thus the presence of such a component is not confirmed, but it is likely. The 
nature of the quiescent emission is still debated.
It could either be due to an incoherent process such, as gyrosynchrotron radiation, or
depolarized electron cyclotron maser instability emission \citep{hallinan08}. On the other hand,
the ECM is confirmed to be the dominant mechanism driving the bulk of the emission
from five of the detected dwarfs \citep{hallinan08, berger+etal09, mclean+etal2011}.
For the L3.5 dwarf 2MASS J00361617+1821104, \citet{hallinan08} showed that the
radio emission is originally due to the ECM, but perhaps depolarized during propagation toward the observer. 
In addition, recent particle-in-cell numerical
simulations by \citet{Yu12} have shown that, under certain circumstances, maser emission can be intrinsically weakly polarized. The same mechanism can be 
applied to generate the bright, 100$\%$ polarized 
burst/flare detected from another ultracool dwarf - DENIS-P J104814.9-395604 \citep{burgasser05}.

Particle-in-cell simulations, which are based on the particle-field theory, indicate that a mildly
relativistic or relativistic electron-beam-driven cyclotron maser can be an effective mechanism to
release electromagnetic waves and heat the surrounding plasmas \citep{Yu12}. Furthermore, from the diffusion process
of the electrons in velocity space, a high-energy tail may be rapidly developed along the direction
perpendicular to the magnetic field, which can eventually evolve to moderately or strongly relativistic
electrons depending on the initial energy of the electron current, and contribute gyrosynchrotron radiation.
This may lead to the appearance of a radio continuum and the deformation of the spectral energy distribution.
Also, the simulations show that a series of discrete spectral lines can appear at certain frequency bands,
which may increase the difficulty of finding the fundamental cyclotron frequency in the observed radio
frequencies.

However, it is likely that the magnetic field inhomogeneity will smooth out these bands, thus producing a continuous
spectrum. Therefore, it is possible that determining the cyclotron frequency and detecting radio emission from a
larger number of UCDs may require observations over a wider frequency band.

In the case of beamed emission, the geometric selection may play an important role in determining the detection rate (see insert 
panel in Fig.~\ref{fig_sptdis}).
The probability $P$ to detect this emission can be expressed as $P=2\theta/(4\pi)$.
The half solid angle of the radiation cone $\theta$ can be approximated by $\cos \theta\approx \frac{\varv_{\rm e}}{c}$, where
$\varv_{\rm e}$ and $c$ are the velocity of electrons in a local region and the speed of light. The current detection
rate for dwarfs over spectral type M7 -- L3.5 is about $\sim 9\%$, which may imply that we need a mildly relativistic
electron beam with an average velocity of at least $0.8c$.

The numerical simulations indicate that a relativistic electron beam can release more than 30$\%$ of its kinetic energy
to electromagnetic energy in a relatively weak magnetic field (hundreds of Gauss) \citep{Yu12}. So one tentative
conclusion is that fast electron beams may be common even in cool objects, e.g., L dwarfs. The electrons may come from
the magnetic field coupled ionization (or discharging) process in the atmosphere, as discussed in \citet{Helling11}.
Alternatively, they can be associated with the internal activity of the ultracool dwarfs. 

The ECM emission and/or geometric selection may be the explanation for the lack of detection of two particular dwarfs in our list
- the M9 dwarf LHS 2924 and the M7 dwarf GJ 3622, both
listed in SIMBAD as flare stars. LHS 2924 has detected H$\alpha$ emission (see Table \ref{tab:table1}), magnetic
field strength of Bf = 1.6 $\pm$ 0.2 kG (where f is the filling factor), and $v~\sin~i$ = 10 km~s$^{-1}$
\citep{reiners+basri07}. Depending on the inclination angle, the rotational velocity could be in the
range 10 to 57 km~s$^{-1}$. Considering the presence of both chromospheric activity and kilogauss fields,
it would not be surprising if radio emission was present as well. However, this is not the case. We
have not detected this dwarf down to a 3$\sigma$ flux level of 6.3~$\times 10^{-5}$~Jy at 4.9 GHz.
\citet{mclean+etal2012} reported a 3$\sigma$ upper limit of 8.4 ~$\times 10^{-5}$~Jy at 8.5 GHz. The
lack of detection at the higher frequency is not surprising considering the magnetic field strength
measured. Detection at 4.9 GHz, on the other hand, should have been possible since the surface magnetic field is not less than 1600 G (i.e., the cyclotron
frequency exceeds 4.5 GHz). Thus, the absence of maser emission may be attributed to geometric effects since
the beamed radio emission might never sweep in the direction of Earth
during a full rotation of the UCD. However, for TVLM 513 the
pulses have a duty cycle of less than 15\%, thus as an alternative explanation we may have missed the range of rotational phase during
which pulsed emission is present.

The second dwarf, GJ 3622, has a detected H$\alpha$ emission (see Table \ref{tab:table1}) and X-ray emission
($\log({\mathrm{L_X}})$ = 25.96 erg~s$^{-1}$), a magnetic field strength of Bf = 600 $\pm$ 200 G, and
$v~\sin~i$ = 3 km~s$^{-1}$ \citep{reiners+basri2010a, lee+etal2010, schmitt04}.  The absence of radio emission from GJ 3622 can be explained by the fact that the
surface magnetic field is only about 600 G, so that the maximal cyclotron frequency
is 1.7 GHz, which is below our observation frequency.

The above considerations, however, cannot explain the absence of
gyrosynchrotron emission, which has a wide directivity pattern and can be
produced at frequencies well above the cyclotron frequency. Therefore, despite a strong magnetic field, the concentration
and/or energy of accelerated electrons in the magnetospheres of the mentioned dwarfs may be insufficient to produce observable radio emission.

\subsection{Rotation}

All radio-detected dwarfs have been found to have high $v \sin i$ ($>$20 km s$^{-1}$) values which
may be interpreted as a dependence on rapid rotation \citep{mclean+etal2012}. However, 
over 50 ultracool dwarfs with measured $v~\sin~i \gtrsim$ 15 km~s$^{-1}$ have not been detected as radio
sources. It seems that fast rotation is a sufficient, but not necessary condition for late-M and L dwarfs
to be radio-active. This tentative conclusion is consistent with the rotation and magnetic activity relation for
a sample of M-dwarfs determined by measuring the equivalent widths of the CaII H and K lines \citep{Browning10}. 

The radio emission from ultracool dwarfs can be strongly associated with their rotation
rate in two ways. First, rotation plays a key role in the stellar dynamo, thus determining
the magnetic field strength. Second, rotation can directly affect the process of particle
acceleration. For the planets of the solar system, the energy of the accelerated particles
can come from three sources: a) interaction of a magnetosphere with the solar wind (this
process occurs at all planets and is especially important for the magnetospheres of the
Earth, Uranus, and Neptune); b) interaction of a magnetosphere with a satellite (like in
the Io-Jupiter system); c) centrifugal acceleration of the magnetospheric plasma.

The latter effect is caused by the fact that in a rigidly rotating magnetosphere,
the centrifugal force exceeds gravity at some distance from the star/planet. This
results in an accumulation of plasma and in an increase of the plasma pressure in
those regions. Eventually, the plasma pressure exceeds the magnetic field
pressure, so that the magnetic field lines become broken. As a result, the
magnetic field at distances exceeding a certain radius does not rotate
with the same rate as the star/planet; current sheets are then formed at the
boundary between the inner (co-rotating) and outer parts of the
magnetosphere. The electric fields in the mentioned current sheets, in turn,
can accelerate electrons up to relativistic energies; thus the rotation
energy is converted into particle energy
\citep{Andre1988,Linsky1992,Usov1992,Leto2006}.

The above process has been proven to be important for the magnetospheres of
Jupiter and (to a lesser degree) Saturn. It is also proposed as the main particle
acceleration mechanism for the radio-emitting star CU Virginis and other similar objects.
Regarding the ultracool dwarfs, if a dwarf is not part of a binary system, the
interaction with an external stellar wind is obviously absent. Satellite-induced radio
emission can be recognized by its specific light curves reflecting both the dwarf's rotation
and the satellite motion \citep{Kuznetsov12}. The satellite-induced particle acceleration
can be ruled out at least for some radio-emitting dwarfs, because this model
cannot account for the strictly periodic emission pulses (which are typical
of some dwarfs) nor for nearly constant quiescent emission (which is
also observed). Thus, for an ultracool dwarf without close-in satellites, one probable
particle acceleration mechanism is the centrifugal acceleration.

Recently, \citet{nichols+etal2012} have estimated the parameters of the electric 
currents produced in the rapidly rotating magnetospheres of ultracool dwarfs 
through angular velocity shear; it has been demonstrated that the power 
carried by these currents can be sufficient to provide the observed intensities 
of radio emission. An assumption in this model is that the spin and magnetic 
axes are co-aligned, which if the polarity were reversed, implies that only 
$\sim$50\% of UCDs would have auroral and radio emission. Coupling this with 
the known radio-active duty cycle of $\sim$ 0.15 phase for UCDs \citep{hallinan08}, 
similar to the active duty cycle of Jupiter of 0.14 phase \citep{higgins+etal96}, could easily imply a very low detection rate. 

Evidently, the efficiency of this mechanism is strongly dependent on the rotation rate:
an increase of the rotation rate both increases the power transferred to the particles
\citep{Schrijver2009} and shifts the acceleration region toward the stellar
surface (for the rotation period of a few hours, the co-rotation radius is
expected to equal just a few stellar radii). We propose the
following explanation of non-detection of radio emission from the dwarfs
LHS 2924 and GJ 3622: although the rotation rate of these dwarfs is
relatively low, it is sufficient to produce the magnetic fields of kilogauss
strength (indeed, the dynamo models predict that the dynamo effects
saturates at relatively low rotation rates corresponding to v about 5 km s$^{-1}$,
so that a further increase in the rotation rate does not affect the magnetic
field strength significantly \citep{mclean+etal2012}). At the same
time, owing to the low rotation rate, the centrifugal acceleration mechanism
at these dwarfs is too weak and cannot provide the necessary density and/or
energy of accelerated electrons.

There is still no conclusive evidence on whether rotation or inclination angle is more important for
detecting radio activity from UCDs. It might be the case that both play a significant role - the first 
for generating the emission and the latter for detecting it. The ultracool dwarfs with
detected ECM emission are all found to be fast rotators with high inclination angles.
The question whether detection of the radio emission from a larger sample depends on observing frequency
may only be addressed via observations of individual targets over wide frequency bands with the
upgraded JVLA and ATCA.

There are currently several ongoing modeling projects that aim at a better understanding
of the emission process that produces the periodic radio bursts with high brightness temperature and
polarization degree in UCDs. Recent computations indicate that the rotation-modulated ECM emission
can interpret the radio light curve from the M8.5 dwarf TVLM 513-46546 \citep{Yu11}, therefore the
properties of the radio emission region can be determined. Future work will focus on the combination
between different magnetic field topology and a particle simulation to reproduce the local
radio emission region.

\citet{Kuznetsov12} presented results for the same dwarf where the emission
properties were similar to the auroral radio emission of the magnetized planets of the solar
system via the electron-cyclotron maser instability. Two models were considered where the
emission was caused by interaction with a satellite or derived from a narrow
sector of active longitudes. It was found that the model of emission from an active sector is able
to qualitatively reproduce the main features of the radio light curves; the magnetic dipole
needs to be highly tilted (by about 60$^{\circ}$) with respect to the rotation axis. For the
most often observationally studied object, TVLM 513-46546, the model of the satellite-induced emission was
inconsistent with the observations.

\subsection{Other possibilities}

Another possible explanation for the lack of detected radio emission could be long-term variability.
To date, there are several ultracool dwarfs whose radio emission varies considerably over long periods, 
with emission levels dropping below the detection limit in some cases \citep{antonova07, berger+etal2010,
mclean+etal2012}.   

Long-term variability, however, requires systematic monitoring of large sample
of targets. Furthermore, full rotational phase coverage should be obtained to
account for the narrow beaming of electron-cyclotron maser emission.

\begin{acknowledgements}
We gratefully acknowledge funding for this project by the Bulgarian National Science Fund
(contract No DDVU02/40/2010). Research at Armagh Observatory is grant-aided
by the N. Ireland Department of Culture, Arts and Leisure. This research has
made use of the SIMBAD database, operated at CDS, Strasbourg, France.
AKU, SYU \& JGD thank the Leverhulme Trust for their support of this project. 
\end{acknowledgements}

\bibliographystyle{aa}

\bibliography{ucd-survey}

\newpage

\appendix
\onecolumn
\section{Activity in H$\alpha$ and X-rays of the sample of radio-observed dwarfs.} \label{appA}

\begin{figure*}
 \centering
 \resizebox{\hsize}{!}{
 \includegraphics[angle=0]{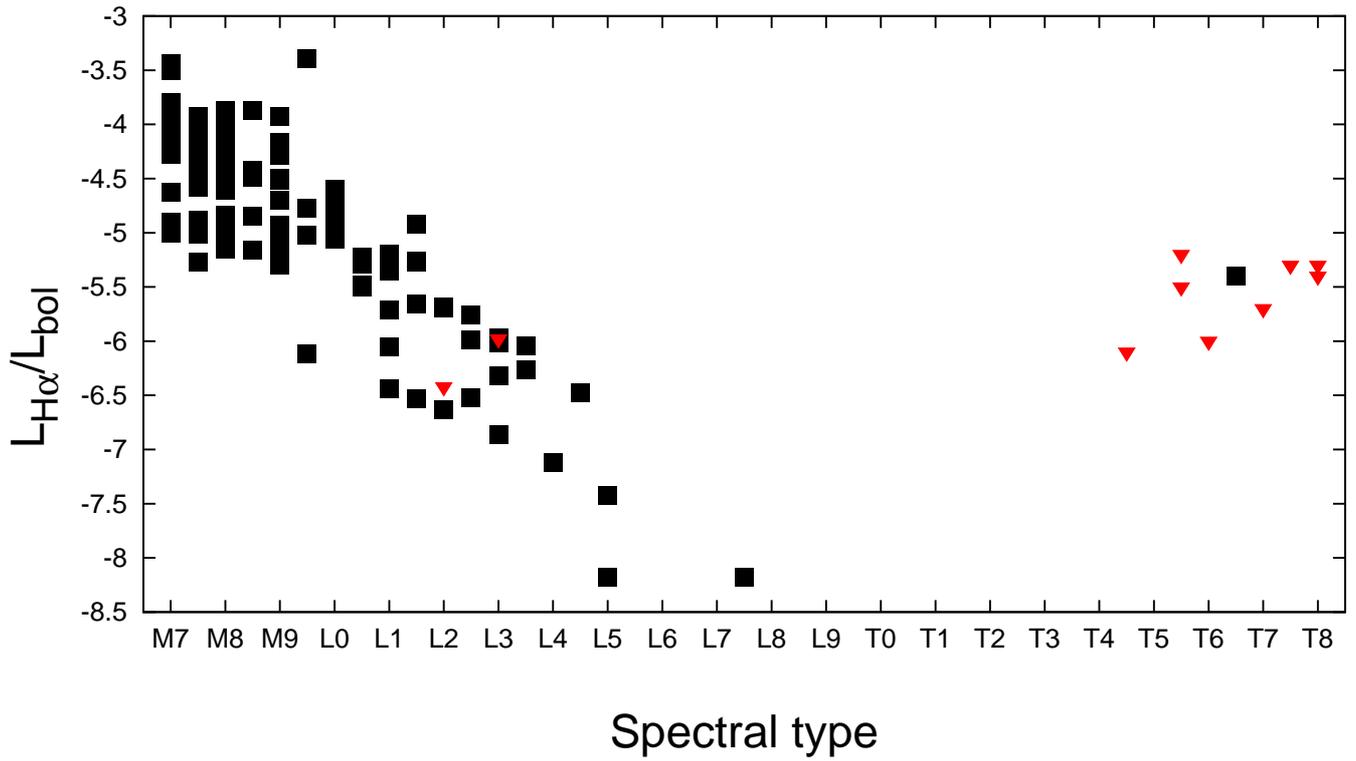}
 }
 \caption[]{H$\alpha$ luminosity - spectral type for all ultracool dwarfs observed at radio frequencies. Squares represent luminosities while triangles show upper limits. Data are taken from the literature \citep[][and references therein]{mclean+etal2012, burgasser+etal03, berger+etal2010}. }  \label{fig:figure_a1}
 \end{figure*}

\begin{figure*}
 \centering
 \resizebox{\hsize}{!}{
 \includegraphics[angle=0]{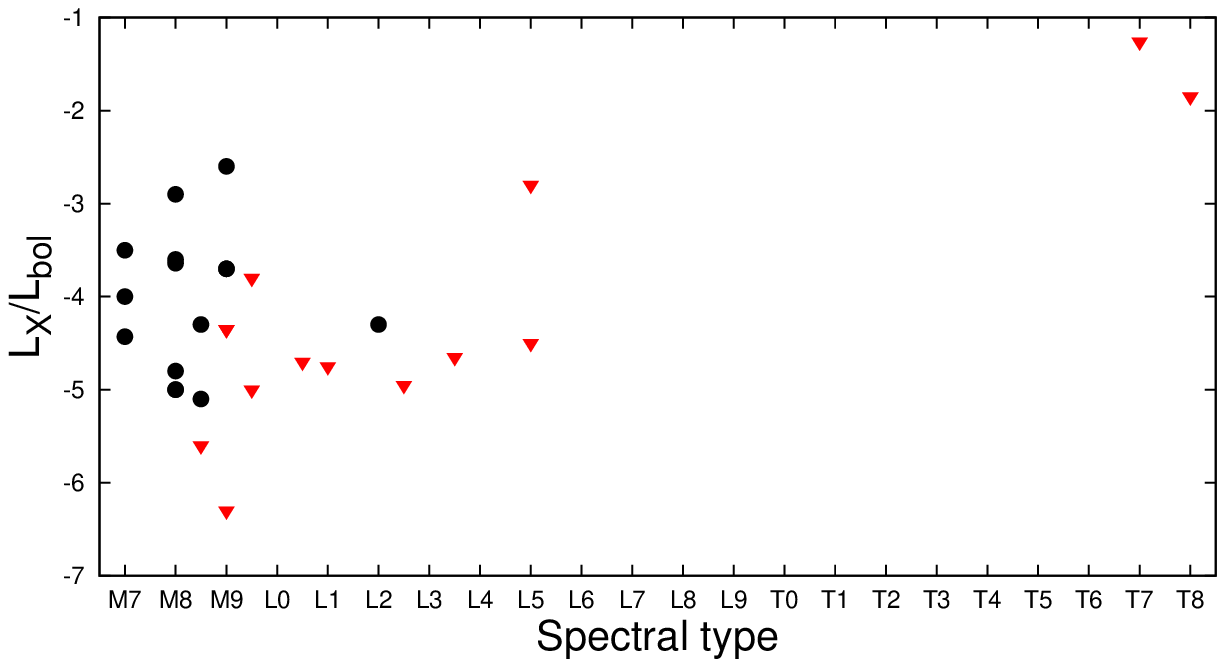}
 }
 \caption[]{X-ray luminosity - spectral type for all ultracool dwarfs observed at radio frequencies. Circles represent luminosities while triangles show upper limits. Data are taken from the literature \citep[][and references therein]{mclean+etal2012, stelzer+etal_2012, berger+etal2010,grosso+etal_07, schmitt04, burgasser+etal03, fleming+etal93}. } \label{figure_a2}
 \end{figure*}

In Table \ref{tab_raducd} we list all ultracool dwarfs observed in the radio domain, their distances, v$\sin$~i, 
radio fluxes, and luminosities/upper limits in H$\alpha$, X-rays, and the radio. About 60$\%$ of the radio-observed 
dwarfs also have observations in H$\alpha$, while only $\approx 18\%$ were observed in X-rays. Figures 
\ref{fig:figure_a1} and \ref{figure_a2} show the H$\alpha$ and X-ray quiescent luminosities as a function of 
spectral type. Despite the trend of lower L$_{H\alpha}$/L$_{bol}$ with later spectral 
type, only three UCDs of type $\leq$ L7.5 have upper limits on their H$\alpha$ luminosities. Toward the later  
spectral types, only one dwarf has a marginal H$\alpha$ detection - the one detected with Arecibo, the T6.5 dwarf 2MASS 
J1047+21 \citep{Route12, burgasser+etal03}. However, its activity level is comparable to that of early-L dwarfs. 
Overall, these results point to chromospheric activity present throughout the L spectral type and possibly as 
far as late-T types.

For late-type, main-sequence active stars (and especially active M dwarfs), there is an empirical correlation 
between the observed X-ray and radio emission ($\mathrm{L_R/L_X \approx 10^{-15.5}}$) found by \citet{gb93}. 
The suggestion is that the same population of relativistic electrons is responsible for both emissions - the 
spiraling motions of the electrons in the magnetic field give rise to incoherent radio emission prior to 
heating the coronal plasma, which leads to thermal X-ray emission. However, for the ultracool dwarfs this 
correlation is found to be violated by several orders of magnitude \citep[][and references there in]{berger+etal2010}. 
For UCDs with observations in both the X-ray and radio domains, \citet{stelzer+etal_2012} suggested that they may  
be separated into two groups: X-ray flaring but radio-faint objects that are also slow rotators, and X-ray-faint 
but strong radio-bursting dwarfs with fast rotation. However, there are several dwarfs with v$\sin i$ in the 
range 15 - 60 km s$^{-1}$ that do not follow this pattern. For example, the L2 dwarf Kelu-1 (with v$\sin i$ = 
60 km s$^{-1}$) has been reported to only possess X-ray emission \citep{audard+etal07}. As seen from Table 
\ref{tab_raducd} and Figure \ref{figure_a2}, the sample of dwarfs observed both in radio and X-rays is too 
small to allow any significant conclusions. \\

%

\begin{center}
\footnotesize
\begin{longtable}{llcccccccc}
\caption{Activity characteristics of all ultracool dwarfs observed at radio frequencies. Data taken from this paper; \citet{mclean+etal2012,berger+etal2010,Route12, stelzer+etal_2012,grosso+etal_07,schmitt04,burgasser+etal03,fleming+etal93} and references therein.}
\label{tab_raducd}\\
\hline
\hline
\multicolumn{1}{l}{Name} & \multicolumn{1}{l}{Other name} & \multicolumn{1}{c}{Sp.T.} &
\multicolumn{1}{c}{$d$} & \multicolumn{1}{c}{$v\sin i$} & \multicolumn{1}{c}{$L_{\rm bol}$} & \multicolumn{1}{c}{$L_{\rm H\alpha}/L_{\rm bol}$}
& \multicolumn{1}{c}{$L_{\rm X}/L_{\rm bol}$} & \multicolumn{1}{c}{$F_{\nu}$} & \multicolumn{1}{c}{L$_{\nu}$}
\\
\multicolumn{1}{l}{(2MASS J)} & \multicolumn{1}{c}{} & \multicolumn{1}{c}{} &
\multicolumn{1}{c}{(pc)} & \multicolumn{1}{c}{(km s$^{-1}$)} & \multicolumn{1}{c}{($L_{\odot}$)} &
\multicolumn{1}{c}{}  & \multicolumn{1}{c}{}  & \multicolumn{1}{c}{$(\mu$Jy)}
 & \multicolumn{1}{c}{(erg s$^{-1}$ Hz$^{-1}$)}
\\
\hline
\endfirsthead

\multicolumn{4}{l}%
{{\bfseries \tablename\ \thetable{} -- continued from previous page}} \\

\hline
\multicolumn{1}{l}{Name} & \multicolumn{1}{l}{Other name} & \multicolumn{1}{c}{Sp.T.} &
\multicolumn{1}{c}{$d$} & \multicolumn{1}{c}{$v\sin i$} & \multicolumn{1}{c}{$L_{\rm bol}$} & \multicolumn{1}{c}{$L_{\rm H\alpha}/L_{\rm bol}$}
& \multicolumn{1}{c}{$L_{\rm X}/L_{\rm bol}$} & \multicolumn{1}{c}{$F_{\nu}$} & \multicolumn{1}{c}{$L_{\nu}$}
\\
\multicolumn{1}{l}{(2MASS J)} & \multicolumn{1}{c}{} & \multicolumn{1}{c}{} &
\multicolumn{1}{c}{(pc)} & \multicolumn{1}{c}{(km s$^{-1}$)} & \multicolumn{1}{c}{($L_{\odot}$)} &
\multicolumn{1}{c}{}  & \multicolumn{1}{c}{}  & \multicolumn{1}{c}{$(\mu$Jy)}
 & \multicolumn{1}{c}{(erg s$^{-1}$ Hz$^{-1}$)}
\\
\hline
\endhead
\endlastfoot
10481258$-$1120082 & GJ 3622 &M7          & 4.5 &3.0& $-$3.16 & $-$4.63  & $-$4.43 &$<$ 63     & $<$ 1.53 $\times 10^{12}$  \\
                   &         &            &   &   & &        & &$<$96      & $<$ 2.33 $\times 10^{12}$  \\
0435161$-$160657 &LP 775$-$31 &M7.0       &9 &... &$-$3.59 &$-$4.28  & ... &$<$48      & $<$ 4.65 $\times 10^{12}$  \\
0440232$-$053008 &LP 655$-$48 &M7.0       &10 &16.5 &$-$3.62 &$-$3.80    & ...&$<$39       & $<$ 4.67 $\times 10^{12}$  \\
0752239+161215 &LP 423$-$31 &M7.0         &11 &9 &$-$3.56 &$-$3.44   & ...&$<$39       & $<$ 5.65 $\times 10^{12}$  \\
1456383$-$280947 &GJ 3877 &M7.0           &7 &8 &$-$3.29 &$-$4.02    & $-$ 4.0 &270$\pm$40  & 1.58 $\times 10^{13}$  \\
1634216+571008 &GJ 630.1B &M7.0           &16 &... &$-$3.13 &...     & ...&$<$530      & $<$ 1.62 $\times 10^{14}$  \\
1655352$-$082340 &VB 8 &M7.0              &6 &9 &$-$3.21 & $-$ 5.0   & $-$ 3.5 &$<$24  & $<$ 1.03 $\times 10^{12}$  \\
                 &       &                &  &  &        & $-$ 2.85(f)   & $-$ 2.85(f) &   &   \\
0741068+173845 &LHS 1937 &M7.0            &18 &10 &$-$3.17 &$-$4.10  &... &$<$75       & $<$ 2.91 $\times 10^{13}$  \\
0818580+233352 &&M7.0                     &19 &4.5 &$-$3.19 &$-$4.11     & ...&$<$78       & $<$ 3.37 $\times 10^{13}$  \\
0952219$-$192431 &&M7.0                   &30 &6 &$-$3.08 &$-$3.94   &... &$<$69       & $<$ 7.43 $\times 10^{13}$  \\
               &&                         &   &  &  &        & &233$\pm$15     & 2.51 $\times 10^{14}$  \\
1141440$-$223215 &&M7.0                   &22 &10 &$-$3.25 &$-$4.90  &... &$<$108      & $<$ 6.25 $\times 10^{13}$  \\
1314203+132001A &&M7.0                    &16 &45 &$-$3.17 &$-$3.97  & ...&1156$\pm$15    & 3.54 $\times 10^{14}$  \\
1354087+084608 &&M7.0                     &17 &... &$-$3.79 &...     &... &$<$105      & $<$ 3.63 $\times 10^{13}$  \\
13564148+4342587 &LP 220$-$13 &M7.0       &16 &14 &$-$3.59 &$-$3.92  & ...&$<$99       & $<$ 3.03 $\times 10^{13}$  \\
1534570$-$141848 &2MUCDa11346 &M7.0       &11 &10 &$-$3.34 &$-$4.01  & ...&$<$87       & $<$ 1.26 $\times 10^{13}$ \\
2337383$-$125027 &LP 763$-$3 &M7.0        &19 &... &$-$2.89 &$-$3.50     &... &$<$84       & $<$ 3.63 $\times 10^{13}$ \\
17571539+7042011 &LP 44$-$162 &M7.5       &12.5& 33 & $-$3.48 & $-$5.01  & ...&$<$ 81      & $<$ 1.51 $\times 10^{13}$    \\
                   &          &           &  &     &     &   & &$<$117     & $<$ 2.19 $\times 10^{13}$  \\
11554286$-$2224586&LP 851$-$346&M7.5      &   9.7&33 &  $-$3.30& $-$4.58 &... & $<$ 66     & $<$ 7.43 $\times 10^{12}$  \\
                  &                       &   &    &   & &$-$4.58        & &$<$90      & $<$ 1.01 $\times 10^{13}$  \\
05395200$-$0059019&SDSS J0539$-$0059&M7.5 &3.84  &  ... &  ...  & ...    & ...&$<$ 60      & $<$ 1.06 $\times 10^{12}$    \\
12505265$-$2121136&DENIS J1250$-$2121&M7.5   & 11 &  ... & $-$3.25 &...   &... & $<$ 66    & $<$ 9.73 $\times 10^{12}$    \\
                     &               &  & &   & &    & &$<$72      & $<$ 1.04 $\times 10^{13}$ \\
0148386$-$302439 &&M7.5                   &18 &48 &$-$3.67 &$-$4.35  & ...&$<$45       & $<$ 1.74 $\times 10^{13}$ \\
0331302$-$304238 &LP 888$-$18 &M7.5       &12 &$<$3 &$-$3.70 &$-$4.07    & ...&$<$72       & $<$ 1.51 $\times 10^{13}$  \\
0417374$-$080000 &&M7.5                   &17 &7 &$-$3.72 &$-$4.32   & ...&$<$36       & $<$ 1.25 $\times 10^{13}$  \\
0429184$-$312356 &&M7.5                   &10 &$<$3 &$-$3.70 &$-$3.93    & ...&$<$48       & $<$ 5.74 $\times 10^{12}$  \\
1521010+505323 &NLTT 40026 &M7.5          &16 &40 &$-$3.70 &$-$4.88  &... &$<$39       & $<$ 1.19 $\times 10^{13}$  \\
0351000$-$005244 &GJ 3252 &M7.5           &15 &6.5 &$-$3.06 &$-$4.16     & ...&$<$123      & $<$ 3.31 $\times 10^{13}$  \\
1006319$-$165326 &LP 789$-$23 &M7.5       &16 &16 &$-$3.28 &$-$4.22  & ...&$<$87       & $<$ 2.67 $\times 10^{13}$  \\
1246517+314811 &LHS 2632 &M7.5            &18 &7.3 &$-$3.25 &$-$5.27     &... &$<$90       & $<$ 3.49 $\times 10^{13}$  \\
1253124+403403 &LP 218$-$8 &M7.5          &17 &9 &$-$3.29 &$-$4.27   & ...&$<$78       & $<$ 2.70 $\times 10^{13}$  \\
1332244$-$044112 &&M7.5                   &21 &9 &$-$3.18 &$-$4.37   & ...&$<$60       & $<$ 3.17 $\times 10^{13}$  \\
1507277$-$200043 &&M7.5                   &14 &64 &$-$3.61 &$-$4.47  & ...&$<$96       & $<$ 2.25 $\times 10^{13}$  \\
1546054+374946 &&M7.5                     &20 &10 &$-$3.25 &$-$3.98  & ...&$<$84       & $<$ 4.02 $\times 10^{13}$  \\
2331217$-$274950 &&M7.5                   &15 &9 &$-$3.06 &$-$4.03   & ...&$<$72       & $<$ 1.94 $\times 10^{13}$  \\
04351455$-$1414468   &  &  M8         & 14 & ... & ... & ...     &... & $<$ 60     & $<$ 1.41 $\times 10^{13}$   \\
02150802$-$3040011&LP885$-$35&  M8         &12.37 &... & $-$3.55 & ...   & ...& $<$ 75     & $<$ 1.37 $\times 10^{13}$   \\
05392474+4038437&LSR J0539+4038&M8         &  10  &  ... &  ...  &...    &... & $<$ 63     & $<$ 7.54 $\times 10^{12}$  \\
0019262+461407 &&M8.0                     &19 &68 &$-$3.80 &$-$4.51  &... &$<$33       & $<$ 1.43 $\times 10^{13}$   \\
0350573+181806 &LP 413$-$53 &M8.0         &23 &4 &$-$3.82 &...       & ...&$<$105      & $<$ 6.65 $\times 10^{13}$  \\
0436103+225956 &&M8.0                     &140 &... &$-$2.62 &...    & ...&$<$45       & $<$ 1.06 $\times 10^{15}$  \\
0517376$-$334902 &&M8.0                   &15 &8 &$-$3.82 &$-$4.42   & ...&$<$54       & $<$ 1.45 $\times 10^{13}$  \\
1016347+275149 &LHS 2243 &M8.0            &16 &$<$3 &$-$3.65 &$-$3.87    & ...&$<$45       & $<$ 1.38 $\times 10^{13}$  \\
1048146$-$395606 &DENIS 1048 &M8.0        &4 &18 &$-$3.39 &$-$5.15   &$-$ 5.0 &140$\pm$40 & 2.68 $\times 10^{12}$  \\
                 &           &            &  &   &    &      &  & 29600$\pm$100 & 5.67 $\times 10^{14}$  \\
1139511$-$315921 &&M8.0                   &20 &... &$-$3.39 &...     &$-$4.8 &$<$99    & $<$ 4.74 $\times 10^{13}$  \\
1534570$-$141848 &&M8.0                   &11 &10 &$-$3.39 &$-$4.01  & ...&$<$111      & $<$ 1.61 $\times 10^{13}$ \\
1843221+404021 &GJ 4073 &M8.0             &14 &5 &$-$3.51 &$-$4.11   &$-$3.64 &$<$48       & $<$ 1.13 $\times 10^{13}$  \\
               &        &                 &   &  &$-$3.09 &$-$4.11   & &$<$96      & $<$ 2.25 $\times 10^{13}$  \\
 \\
1916576+050902 &VB 10 &M8.0               &6 &6.5 &$-$3.35 &$-$4.9   & $-$ 5.0 &$<$81   & $<$ 3.49 $\times 10^{12}$  \\
               &      &                   &   &   &         &$-$4.4(f)            & $-$ 4.1(f) &   & \\
2037071$-$113756 &&M8.0                   &17 &$<$3 &$-$3.74 &$-$5.02    & ...&$<$33       & $<$ 1.14 $\times 10^{13}$  \\
0027559+221932 &LP 349$-$25 B &M8.0       &10 &56 &$-$3.12 &$-$4.53  &... &323$\pm$14     & 3.86 $\times 10^{13}$  \\
0248410$-$165121 &&M8.0                   &17 &$<$3 &$-$3.45 &$-$4.25    & ...&$<$81       & $<$ 2.80 $\times 10^{13}$  \\
0320596+185423 &LP 412$-$31 &M8.0         &15 &15 &$-$3.26 &$-$3.87  & $-$ 3.6 &$<$81      & $<$ 2.18 $\times 10^{13}$  \\
        &       &     & & &       &          & $-$ 0.65(f) &   &   \\
0544115$-$243301 &&M8.0                   &19 &$<$3 &$-$3.33 &$-$4.12    & ...&$<$63       & $<$ 2.72 $\times 10^{13}$  \\
0629235$-$024851B &GJ 234 B &M8.0         &4 &... &$-$3.00 &...      & $-$ 2.9&$<$81       & $<$ 1.55 $\times 10^{12}$  \\
1016347+275149 &LHSa2243 &M8.0            &14 &$<$3 &$-$3.38 &$-$3.87    & ...&$<$84       & $<$ 1.97 $\times 10^{13}$  \\
1024099+181553 &2MUCDa10906 &M8.0         &16 &5 &$-$3.38 &$-$4.84   &...&$<$87    & $<$ 2.67 $\times 10^{13}$  \\
1309218$-$233035 &&M8.0           &16 &7 &$-$3.63 &$-$4.35       & ...&$<$93       & $<$ 2.85 $\times 10^{13}$  \\
1428041+135613 &LHS 2919 &M8.0            &10 &... &$-$3.37 &...     & ...&$<$90       & $<$ 1.08 $\times 10^{13}$  \\
1440229+133923 &&M8.0                     &18 &$<$3 &$-$3.33 &$-$4.60    &... &$<$75       & $<$ 2.91 $\times 10^{13}$  \\
1444171+300214 &LP 326$-$21 &M8.0         &13 &... &$-$3.61 &...     & ...&$<$81       & $<$ 1.64 $\times 10^{13}$  \\
2206227$-$204706 &&M8.0           &27 &24 &$-$2.95 &$-$4.54      &... &$<$84       & $<$ 7.33 $\times 10^{13}$  \\
2349489+122438 &LP 523$-$55 &M8.0         &20 &4 &$-$3.31 &$-$4.61   &... &$<$60       & $<$ 2.87 $\times 10^{13}$  \\
2351504$-$253736A &&M8.0          &18 &36 &$-$3.36 &$-$4.61      & ...&$<$69       & $<$ 2.67 $\times 10^{13}$  \\
18261131+3014201&LSR J1826+3014&M8.5      &13.9 &... &  ... & ...    &... & $<$ 87     & $<$ 2.01 $\times 10^{13}$  \\
0335020+234235 &&M8.5                     &19 &... &$-$3.61 &...     &... &$<$69       & $<$ 2.98 $\times 10^{13}$  \\
1454290+160605 &GJ 569Ba &M8.5            &10 &... &$-$3.80 &...     & $-$4.3&$<$30    & $<$ 3.59 $\times 10^{12}$  \\
1501081+225002 &TVLM513$-$465 &M8.5       &11 &60 &$-$3.59 &...      & $-$5.1 &190$\pm$15     & 2.75 $\times 10^{13}$  \\
               &              &           &   &   &    &         & &980$\pm$40     & 1.42 $\times 10^{14}$  \\
               &              &           &   &   &    &         & &4000       & 5.79 $\times 10^{14}$  \\
1835379+325954 &LSR J1835+3 &M8.5         &6 &44 &$-$3.93 &$-$4.85   & $< -$5.6 &525$\pm$15     & 2.26 $\times 10^{13}$ \\
0140026+270150 &&M8.5                     &19 &6.5 &$-$3.32 &...     & ...&$<$20       & $<$ 8.64 $\times 10^{12}$  \\
1121492$-$131308 &GJ 3655 &M8.5           &12 &27 &$-$3.68 &$-$3.87  & ...&$<$102      & $<$ 1.76 $\times 10^{13}$  \\
1124048+380805 &&M8.5                     &19 &7.5 &$-$3.41 &$-$5.16     & ...&$<$66       & $<$ 2.85 $\times 10^{13}$  \\
1403223+300754 &&M8.5                     &19 &10 &$-$3.39 &$-$4.49  &... &$<$60       & $<$ 2.59 $\times 10^{13}$  \\
2353594$-$083331 &&M8.5                   &22 &4.5 &$-$3.41 &$-$4.42     & ...&$<$69       & $<$ 3.99 $\times 10^{13}$  \\
14284323+3310391   &  GJ 3849& M9     &11 &10 & $-$3.62 &$-$4.7  & $< -$4.35& $<$ 63       & $<$ 9.12 $\times 10^{12}$  \\
0019457+521317 &&M9.0             &19 &9 &$-$3.95 &$-$4.29       & ...&$<$42       & $<$ 1.81 $\times 10^{13}$  \\
0109511$-$034326 &LP 647$-$13 &M9.0       &11 &13 &$-$3.98 &$-$4.50  & ...&$<$33       & $<$ 4.78 $\times 10^{12}$  \\
0339352$-$352544 &LP 944$-$20 &M9.0   &5 &26 &$-$3.79 &$-$5.30   & $< -$6.3 &74$\pm$13   & 2.21 $\times 10^{12}$  \\
               &          &               &  &   &  &        &$-$3.7(f) & 2600$\pm$200  & 7.78 $\times 10^{13}$  \\
0434152+225031 &&M9.0                     &140 &... &$-$2.53 &...    & &$<$69      & $<$ 1.62 $\times 10^{15}$  \\
0436389+225812 &&M9.0                     &140 &... &$-$2.59 &...    &$-$3.7 &$<$57    & $<$ 1.34 $\times 10^{15}$  \\
0537259$-$023432 &&M9.0                   &352 &... &$-$3.56 &...    & ...&$<$66       & $<$ 9.78 $\times 10^{15}$  \\
0810586+142039 &&M9.0                     &20 &11 &$-$3.39 &$-$4.17      & ...&$<$39       & $<$ 1.87 $\times 10^{13}$  \\
0853362$-$032932 &GJ 3517 &M9.0           &9 &13.5 &$-$3.49 &$-$3.93     & $-$3.7 &$<$81   & $<$ 7.85 $\times 10^{12}$  \\
                 &  &        & &    &         &           & $-$2.5(f)       &   & \\
1454280+160605 &GJ 569Bb &M9.0            &10 &... &$-$4.04 &...     &$-$2.6 &$<$30    & $<$ 3.59 $\times 10^{12}$  \\
1627279+810507 &&M9.0                     &21 &... &$-$3.45 &...     &... &$<$60       & $<$ 3.17 $\times 10^{13}$  \\
1707183+643933 &&M9.0                     &17 &... &$-$3.44 &...     & ...&$<$60       & $<$ 2.07 $\times 10^{13}$  \\
0443376+000205 &&M9.0                     &16 &13.5 &$-$3.47 &$-$5.00    & ...&$<$54       & $<$ 1.65 $\times 10^{13}$  \\
1224522$-$123835 &&M9.0                   &17 &7 &$-$3.94 &$-$4.52   &... &$<$102      & $<$ 3.53 $\times 10^{13}$  \\
1411213$-$211950 &&M9.0                   &16 &44 &$-$3.93 &$-$4.93  &... &$<$93       & $<$ 2.85 $\times 10^{13}$  \\
1428432+331039 &LHS 2924 &M9.0            &11 &11 &$-$3.59 &$-$5.14  & $< -$4.35 &$<$84   & $<$ 1.22 $\times 10^{13}$  \\
1707234$-$055824 &2MUCD 20701 &M9.0       &15 &... &$-$3.31 &...     &... &$<$81       & $<$ 2.80 $\times 10^{13}$  \\
          &&                          & &   & & &            &$<$48    & $<$ 1.29 $\times 10^{13}$  \\
2200020$-$303832AB & &M9.0 &35 &17 &$-$3.17 &$-$5.03     & ...&$<$78       & $<$ 1.14 $\times 10^{14}$  \\
0024246$-$015819 &BRI B0021$-$0&M9.5      &12 &33 &$-$3.50 &$-$6.12  &$< -$5.0 &83$\pm$18      & 1.43 $\times 10^{13}$  \\
                 &              &         &   &   &$-$3.45 &$-$6.12  & &$<$60      & $<$ 1.03 $\times 10^{13}$  \\
0027420+050341 &PC 0025+044 &M9.5         &72 &13 &$-$3.62 &$-$3.39  & $< -$3.8&$<$75      & $<$ 4.65 $\times 10^{14}$  \\
0109217+294925 &&M9.5                     &19 &7 &$-$3.49 &...       & ...&$<$54       & $<$ 2.33 $\times 10^{13}$  \\
0149089+295613 &&M9.5                     &17 &12 &$-$3.74 &...      & ...&$<$140      & $<$ 4.84 $\times 10^{13}$  \\
1438082+640836 &&M9.5                     &18 &12 &$-$4.08 &$-$4.77  &... &$<$105      & $<$ 4.07 $\times 10^{13}$  \\
2237325+392239 &G216$-$7B &M9.5           &19 &... &$-$3.66 &$-$5.02     & ...&$<$81       & $<$ 3.50 $\times 10^{13}$  \\
\hline
17312974+2721233& LSPM J1731+2721&L0      & 11.8&15& $-$3.74 & $-$4.6    & ...& $<$ 57     &  $<$ 9.50 $\times 10^{12}$ \\
1421314+182740 &&L0.0                     &20 &... &$-$3.56 &...     & ...&$<$42       &  $<$ 2.01 $\times 10^{13}$ \\
0345431+254023 &&L0.0                 &27 &... &$-$3.56 &...     & ...&$<$87       &  $<$ 7.59 $\times 10^{13}$ \\
0314034+160305 &&L0.0                 &14 &19 &$-$3.59 &$-$4.69  & ...&$<$108      &  $<$ 2.53 $\times 10^{13}$ \\
1159385+005726 &&L0.0                     &30 &71 &$-$3.57 &$-$5.06  & ...&$<$54       &  $<$ 5.81 $\times 10^{13}$ \\
1221277+025719 &&L0.0                 &19 &25 &$-$3.59 &$-$4.88  & ...&$<$78       &  $<$ 3.37 $\times 10^{13}$ \\
1731297+272123 &&L0.0                 &12 &15 &$-$3.56 &$-$4.80  &... &$<$69       &  $<$ 1.19 $\times 10^{13}$ \\
1854459+842947 &&L0.0                 &23 &7 &$-$3.62 &$-$4.73   & ...&$<$87       &  $<$ 5.51 $\times 10^{13}$ \\
0746425+200032 &&L0.5                 &12 &31 &$-$3.93 &$-$5.29  &$< -$4.7 &224$\pm$15     & 3.86 $\times 10^{13}$ \\
               &&                         &   &   &  &       & &15000$\pm$100  & 2.58 $\times 10^{15}$  \\
1412244+163311 &&L0.5                 &25 &19 &$-$3.61 &$-$5.50  & ...&$<$69       & $<$ 5.16 $\times 10^{13}$  \\
1441371$-$094559 &&L0.5                   &28 &23 &$-$3.59 &$-$5.48  & ...&$<$84       & $<$ 7.88 $\times 10^{13}$  \\
2351504$-$253736B &&L0.5              &18 &41 &$-$3.36 &$-$5.22  & ... &$<$69      & $<$ 2.68 $\times 10^{13}$  \\
0602304+391059 &LSR 0602+39 &L1.0         &11 &9 &$-$4.28 &$-$6.05   & $< -$4.75&$<$30     & $<$ 4.34 $\times 10^{13}$  \\
1300425+191235 &&L1.0                 &14 &10 &$-$4.12 &$-$5.71  & ...&$<$87       & $<$ 2.04 $\times 10^{13}$  \\
0235599$-$233120 &&L1.0               &21 &13 &$-$3.63 &$-$6.44  & ...&$<$99       & $<$ 5.22 $\times 10^{13}$  \\
1045240$-$014957 &&L1.0               &17 &$<$3 &$-$3.65 &$-$6.44    &... &$<$57       & $<$ 1.97 $\times 10^{13}$  \\
1048428+011158 &&L1.0             &15 &17 &$-$3.69 &$-$5.71  & ...&$<$21       & $<$ 8.08 $\times 10^{13}$  \\
1439283+192914 &&L1.0             &14 &11 &$-$3.67 &$-$5.20  &... &$<$78       & $<$ 1.83 $\times 10^{13}$  \\
1555157$-$095605 &&L1.0               &13 &11 &$-$3.68 &$-$5.35  & ...&$<$84       & $<$ 1.70 $\times 10^{13}$  \\
0213288+444445 &&L1.5                 &19 &... &$-$4.24 &...     & ...&$<$30       & $<$ 1.30 $\times 10^{13}$  \\
2057540$-$025230 &&L1.5               &16 &62 &$-$4.23 &$-$4.92  & ...&$<$36       & $<$ 1.10 $\times 10^{13}$  \\
1145571+231729 &GL Leo         &L1.5 &44&14 &$-$3.70&$-$5.27        & ...&$<$90    & $<$ 2.08 $\times 10^{14}$  \\
1334062+194035 &&L1.5             &46 &30 &$-$3.68 &$-$6.53  & ...&$<$60       & $<$ 1.52 $\times 10^{14}$  \\
1645221$-$131951 &&L1.5               &12 &9 &$-$3.69 &$-$5.66   & ...&$<$108      & $<$ 1.86 $\times 10^{13}$  \\
1807159+501531 &2MUCDa11756 &L1.5         &15 &76 &$-$3.71 &$-$5.26  &... &$<$84       & $<$ 2.26 $\times 10^{13}$  \\
               &            &         &   &   &$-$4.24 &$-$5.26  &  &$<$39     & $<$ 1.05 $\times 10^{13}$  \\
09211410$-$2104446&DENIS J0921$-$2104 &L2.0 &12 &15 &$-$4.01 &$<$ $-$6.42  &... & $<$ 72       & $<$ 1.14 $\times 10^{13}$   \\
                  &                   &   &   & &$-$3.83 &$<$ $-$6.42    & &$<$75      & $<$ 1.29 $\times 10^{13}$  \\
08283419$-$1309198&DENIS J0828$-$1309&L2.0  &11.6 &33&$-$3.64& $-$5.68   & ...& $<$ 63     & $<$ 1.01 $\times 10^{13}$  \\
                     &                    &   &  &  &       &$-$6.63   &  &$<$66       & $<$ 1.06 $\times 10^{13}$  \\
0109015$-$510049 &&L2.0           &10 &... &$-$3.89 &...     & ...&$<$111      & $<$ 1.33 $\times 10^{13}$  \\
0445538$-$304820 &&L2.0           &17 &... &$-$4.33 &...     & ...&$<$66       & $<$ 2.28 $\times 10^{13}$   \\
1305401$-$254110 &Kelu$-$1 &L2.0      &19 &60 &$-$3.57 &$-$5.69  & $-$ 4.3& $<$27  & $<$ 1.17 $\times 10^{13}$  \\

0523382$-$140302 &&L2.5           &13 &21 &$-$4.39 &$-$6.52  &$< -$4.95  &$<$39   & $<$ 7.89 $\times 10^{12}$  \\
               &&                 &   &   &  &       &    &231$\pm$14     & 4.67 $\times 10^{13}$  \\
1029216+162652 &&L2.5             &23 &29 &$-$3.72 &$-$5.76  &... &$<$33       & $<$ 2.09 $\times 10^{13}$  \\
1047310$-$181557 &&L2.5           &22 &15 &$-$3.86 &$-$5.99  & ...&$<$63       & $<$ 3.65 $\times 10^{13}$  \\
0251149$-$035245 &&L3.0           &12 &... &$-$4.42 &...     &...&$<$36    & $<$ 6.20 $\times 10^{12}$  \\
1721039+334416 &&L3.0             &15 &... &$-$4.46 &...     & ...&$<$48       & $<$ 1.29 $\times 10^{13}$  \\
2104149$-$103736 &&L3.0           &17 &27 &$-$4.47 &$-$5.97  &... &$<$24       & $<$ 8.30 $\times 10^{12}$  \\
0913032+184150 &&L3.0             &46 &34 &$-$3.77 &$-$6.86  &... &$<$102      & $<$ 2.58 $\times 10^{14}$  \\
1203581+001550 &&L3.0             &19 &39 &$-$3.84 &$-$6.02  &... &$<$63       & $<$ 2.72 $\times 10^{13}$  \\
1506544+132106 &&L3.0             &14 &20 &$-$3.80 &$-$6.32  &... &$<$78       & $<$ 1.83 $\times 10^{13}$  \\
1615441+355900 &&L3.0             &24 &13 &$-$3.82 &$<$ $-$5.98  &... &$<$75       & $<$ 5.17 $\times 10^{13}$  \\
1707234$-$055824 &2MUCD 20701 &L3.0       &17 &... &$-$3.80 &...     & ...&$<$81       & $<$ 2.80 $\times 10^{13}$  \\
0700366+315726A     &    &  L3.5     & 12.2 & 41&$-$3.96&  ...   &... & $<$ 42     & $<$ 8.01 $\times 10^{12}$  \\
                   &        &             &    & &$-$3.88 &$-$6.04   &  &$<$78     &  $<$ 1.39 $\times 10^{13}$ \\
0036161+182110 &&L3.5             &9 &45 &$-$4.51 &$-$6.26   & $< -$4.65 &134$\pm$16     &  1.30 $\times 10^{13}$ \\
               &&                 &  &   &  &        &  &720$\pm$40     &  6.98 $\times 10^{13}$  \\
0045214+163444 &&L3.5             &10 &... &$-$4.58 &...     &... &$<$39       & $<$ 4.67 $\times 10^{12}$  \\
05002100+0330501   &   &  L4.0    & 13.03 &... &$-$4.26  & ...   &... & $<$ 51     & $<$ 1.04 $\times 10^{13}$  \\
1424390+091710 &&L4.0             &32 &... &$-$4.04 &...     &... &$<$96       & $<$ 1.18 $\times 10^{14}$  \\
1705483$-$051646 &&L4.0           &11 &26 &$-$4.65 &$-$7.12  & ...&$<$45       & $<$ 6.51 $\times 10^{12}$  \\
04351455$-$1414468   &     &  L4.5    &9.8 & ... &$-$4.12  & ...     &... & $<$ 42     & $<$ 4.83 $\times 10^{12}$  \\
0141032+180450 &&L4.5             &13 &... &$-$4.76 &...     &... &$<$30       & $<$ 6.07 $\times 10^{12}$  \\
0652307+471034 &&L4.5             &11 &... &$-$4.66 &...     &... &$<$33       & $<$ 4.78 $\times 10^{12}$  \\
2224438$-$015852 &&L4.5           &11 &32 &$-$4.76 &$-$6.48  & ...&$<$33       & $<$ 4.78 $\times 10^{12}$  \\
03552337+1133437   &        &  L5.0     &12.6 &10 &$-$4.03 & ...     & ...& $<$ 45     & $<$ 8.55 $\times 10^{12}$  \\
05395200$-$0059019 &SDSS J0539$-$0059 &L5.0  &13.1 &34 &$-$4.2 & ...     &... & $<$ 48     & $<$ 9.86 $\times 10^{12}$  \\
0004348$-$404405 &GJ 1001BC &L5.0     &10 &42 &$-$4.67 &$-$7.42  &... &$<$45       & $<$ 5.38 $\times 10^{12}$  \\
0144353$-$071614 &&L5.0           &13 &... &$-$4.73 &...     & ...&$<$33       & $<$ 6.67 $\times 10^{12}$  \\
0205034+125142 &&L5.0             &27 &... &$-$4.67 &...     &...&$<$48    & $<$ 4.19 $\times 10^{13}$  \\
0835425$-$081923 &&L5.0           &9 &23 &$-$4.60 &$-$7.42   & ...&$<$30       & $<$ 2.91 $\times 10^{12}$  \\
1228152$-$154734 &&L5.0           &20 &22 &$-$4.19 &...      &$< -$2.8&$<$87   & $<$ 4.16 $\times 10^{13}$  \\
1507476$-$162738 &&L5.0           &7 &32 &$-$4.23 &$-$8.18   &$< -$4.5 &$<$57   & $<$ 3.34 $\times 10^{12}$  \\
17502484$-$0016151   &     &  L5.5      &8 & ... & ...  &  ...   & ...& $<$ 43     & $<$ 4.82 $\times 10^{12}$  \\
1515008+484741 &&L6.0              &9 &... &$-$5.11 &...     &...&$<$27    & $<$ 2.62 $\times 10^{12}$  \\
21481633+4003594   &     &  L6.5       & 7.9 & ... &$-$4.07 & ...    & ...& $<$ 63     & $<$ 4.70 $\times 10^{12}$  \\
0700366+315726B     &     &  L6.0     & 12.2 & 41&$-$3.96&  ...  & ...& $<$ 42     & $<$ 8.01 $\times 10^{12}$   \\
0439010$-$235308 &&L6.5                &11  &... &$-$5.10 &...   & ... &$<$42      & $<$ 6.08 $\times 10^{12}$  \\
09083803+5032088   & ...         &  L7.0     & 15.9 & 31 & ... & ...     & ...& $<$ 111    & $<$ 3.36 $\times 10^{13}$  \\
0030300$-$145033 &&L7.0            &27 &... &$-$5.01 &...    &... &$<$57       & $<$ 4.97 $\times 10^{13}$  \\
0205294$-$115929 &&L7.0            &20 &22 &$-$4.65 &...     & ...&$<$30       & $<$ 1.44 $\times 10^{13}$  \\
1728114+394859 &&L7.0              &24 &... &$-$4.86 &...    & ...&$<$54       & $<$ 3.72 $\times 10^{13}$  \\
0423485$-$041403 &&L7.5            &15 &... &$-$4.83 &...    & ...&$<$42       & $<$ 1.13 $\times 10^{13}$  \\
0825196+211552 &&L7.5              &11 &19 &$-$5.21 &$-$8.18     & ...&$<$45       & $<$ 6.51 $\times 10^{12}$  \\
2252107$-$173013 &&L7.5            &8 &... &$-$5.29 &...     &... &$<$30       & $<$ 2.30 $\times 10^{12}$  \\
08575849+5708514   &SDSS J0857+5708 &L8.0    & 11 & ... & ...&  ...  & ...& $<$ 51     & $<$ 7.38 $\times 10^{12}$  \\
02572581$-$3105523&DENIS J0257$-$3105&L8.0   &9.6 & ... &$-$4.82 & ...   &... &$<$ 93      & $<$ 1.03 $\times 10^{13}$  \\
0929336+342952 &&L8.0              &22 &... &$-$5.25 &...    &... &$<$42       & $<$ 2.43 $\times 10^{13}$  \\
1523226+301456 &&L8.0              &19 &... &$-$5.27 &...    &... &$<$45       & $<$ 1.93 $\times 10^{13}$  \\
1632291+190441 &&L8.0              &15 &30 &$-$5.31 &...     &... &$<$54       & $<$ 1.45 $\times 10^{13}$  \\
08300825+4828482 &SDSS J0830+4828 &L9.0    &13 & ... &$-$4.58 &...   &... & $<$ 87     & $<$ 1.78 $\times 10^{13}$  \\
\hline
0151415+124430 &&T0.0             &21 &... &$-$5.37 &...     &... &$<$51       & $<$ 2.69 $\times 10^{13}$  \\
2204105$-$564657A &&T1.0          &4 &... &$-$5.03 &...      &... &$<$72       & $<$ 1.38 $\times 10^{12}$  \\
01365662+0933473 &SIMP J0136+0933 &  T2.5  &6.4 & ... & ...& ...     & ...& $<$ 75     & $<$ 3.68 $\times 10^{13}$  \\
               &          &       &   8.27  &     &  &       & &           &  $<$ 6.14 $\times 10^{13}$ \\
0207428+000056 &&T4.0             &29 &... &$-$5.21 &...     &... &$<$39       & $<$ 3.92 $\times 10^{13}$  \\
0559191$-$140448 &&T4.5           &10 &... &$-$4.53 &$< -$6.1    &... &$<$27       & $<$ 3.23 $\times 10^{12}$  \\
15031961+2525196   &         &  T5.5    &7 &36 & ... & $< -$5.5  &... & $<$ 90     & $<$ 5.28 $\times 10^{12}$  \\
1534498$-$295227 &&T5.5           &14 &... &$-$5.00 &$< -$5.2    &... &$<$63       & $<$ 1.48 $\times 10^{13}$  \\
09373487+2931409   &       &  T6.0       &6.14 & ... &$-$5.28 &$< -$6.0 &... & $<$ 66      & $<$ 2.98 $\times 10^{12}$  \\
21543318+5942187   &       &  T6.0     &18.8  &  ... & ...  &  ...   & ... & $<$ 60    & $<$ 2.54 $\times 10^{13}$  \\
1624144+002916 &&T6.0                     &11 &... &$-$5.16 &...     & ... &$<$36      & $<$ 5.21 $\times 10^{12}$  \\
2204105$-$564657B &&T6.0          &4 &... &$-$5.03 &...      & ... &$<$72      & $<$ 1.34 $\times 10^{12}$  \\
1047539+212423 &&T6.5             &11 &... &$-$5.35 & $-$5.4     & ...&$<$45       & $<$ 6.51 $\times 10^{12}$  \\
               & &                        &   &    &        &        &  &$\sim$1500     & $\sim $2.17 $\times 10^{14}$  \\
1346464$-$003150 &&T6.5           &15 &... &$-$5.00 &...     & ...&$<$105      & $<$ 2.83 $\times 10^{13}$  \\
07271824+1710012   &         &  T7.0      &9.09&...&$-$5.26&$<$ $-$5.7  &... &$<$ 54       & $<$ 5.34 $\times 10^{12}$  \\
0610351$-$215117 &GJ 229B &T7.0        &6 &... &$-$5.21 &...     & $< -$1.26 &$<$69   & $<$ 2.97 $\times 10^{12}$  \\
11145133$-$2618235   &      &  T7.5   &7  & ... &... & ...   & ...& $<$ 60     & $<$ 3.52 $\times 10^{12}$  \\
               &          &            &   10  &  &  &       &  &          & $<$ 7.18 $\times 10^{12}$ \\
1217111$-$031113 &&T7.5           &11 &... &$-$5.32 &$< -$5.3    &...&$<$111       & $<$ 1.61 $\times 10^{13}$  \\
09393548$-$2448279   &       &  T8.0    &8.7 &... &$-$5.69  & ...    & ... & $<$ 54    & $<$ 4.89 $\times 10^{12}$  \\
           &        &             &5.34  &     &    &    &     &       & $<$ 1.84 $\times 10^{12}$         \\
14571496$-$2121477   &  GL570D    &  T8.0   &5.88  &30 & $-$5.52 &$< -$5.3 &$< -$ 1.85 & $<$ 54    & $<$ 2.23 $\times 10^{12}$  \\
0415195$-$093506 &&T8.0           &6 &... &$-$5.73 &$< -$5.4 & ... &$<$45      & $<$ 1.94 $\times 10^{12}$  \\
\hline
\hline
\end{longtable}
\end{center}


\end{document}